\documentclass[a4paper,11pt]{article}
\usepackage{amsmath, amssymb, graphics, graphicx}
\usepackage{color}
\usepackage{physics}
\usepackage{longtable}
\usepackage{subfigure}
\usepackage{multirow}
\usepackage{slashed}
\usepackage{mathtools}
\usepackage{stackengine}
\stackMath

\newcommand{\mathsym}[1]{{}}

\newcommand{\rmd}{\mathrm{d}}

\allowdisplaybreaks

\usepackage[shortlabels]{enumitem}

\begin{document}


\begin{titlepage}
\begin{center}
\vspace*{-1.0cm}
\hfill {\footnotesize 23-12-2024}

\vspace{2.0cm}

{\LARGE  {\fontfamily{lmodern}\selectfont \bf A Conformal Approach to Carroll  Gravity}} \\[.2cm]

\vskip 1.5cm
{\bf Eric A. Bergshoeff$^{\,a}$, Patrick Concha$^{b,c}$, Octavio Fierro$^{d}$, \\[.1truecm]
Evelyn Rodr\'\i guez$^{b,c}$ and Jan Rosseel$^e$}\\
\vskip 1.2cm

\begin{small}
{}$^a$ \textit{Van Swinderen Institute, University of Groningen \\
Nijenborgh 3, 9747 AG Groningen, The Netherlands} \\
\vspace{1mm}

\vspace{5mm}

{}$^b$ \textit{Departamento de Matem\'atica y F\'\i sica Aplicadas, \\
Universidad Cat\'olica de la Sant\'\i sima Concepci\'on,\\
Alonso de Ribera 2850, Concepci\'on, Chile} \\

\vspace{5mm}

{}$^c$ \textit{Grupo de Investigación en Física Teórica, GIFT, \\
	Universidad Cat\'olica de la Sant\'\i sima Concepci\'on,\\
Alonso de Ribera 2850, Concepci\'on, Chile} \\

\vspace{5mm}

{}$^d$ \textit{Facultad de Ingenier\'{i}a, Arquitectura y Diseño, Universidad San Sebasti\'{a}n,\\
Lientur 1457, Concepción 4080871, Chile}\\

\vspace{5mm}

{}$^e$ \textit{Division of Theoretical Physics, Rudjer Bo\v{s}kovi\'c Institute, \\
Bijeni\v{c}ka 54, 10000 Zagreb, Croatia} \\
\vspace{2mm}
{\texttt{e.a.bergshoeff[at]rug.nl, patrick.concha[at]ucsc.cl, ofierro27[at]yahoo.com, erodriguez[at]ucsc.cl,
Jan.Rosseel[at]irb.hr}}

\end{small}

\end{center}

\vskip .5cm
\begin{abstract}
\vskip.1cm\noindent

We show how to take the first step in the conformal program for constructing general matter couplings to Carroll gravity. In particular, we couple a single massless  electric/magnetic scalar to conformal Carroll gravity with isotropic dilatations and show how, upon gauge-fixing, we obtain a (non-conformal version of) electric/magnetic Carroll gravity. We determine the full Carroll transformation rules paying  special attention to the way  the so-called intrinsic torsion tensors occur in these transformation rules. A noteworthy feature in the magnetic case is that the Lagrange multiplier present in the Lagrangian  gets absorbed, after coupling to conformal Carroll gravity and gauge-fixing, into one of the independent spin-connections of magnetic Carroll gravity. Our results form a convenient starting point for constructing general matter couplings to Carroll gravity. Surprisingly, we find that the same relation between dynamical matter and gravity, which forms the basis of the conformal program, does not work in the usual way in the Galilei case.

\end{abstract}

\end{titlepage}



%

\section{Introduction}

Carroll symmetries have attracted a lot of attention in recent years due to a variety of reasons. First of all, they are the natural symmetries of any null surface. For  this reason they play a central role in investigations of black hole horizons \cite{Donnay:2019jiz} and the null infinity of flat spacetime. Moreover, a conformal extension of Carroll symmetries can be identified with the asymptotic BMS symmetries of flat spacetime that play a fundamental role in Carroll holography \cite{Duval:2014uva,Bagchi:2023cen,Donnay:2023mrd}. Carroll gravity theories have also been constructed. They occur in two versions called electric Carroll gravity \cite{Henneaux:1979vn} and magnetic Carroll gravity \cite{Bergshoeff:2017btm}.

It is natural to extend Carroll gravity to include  matter couplings. One way to obtain such matter couplings is to start from a known relativistic system with matter coupled to gravity and perform a Carroll expansion \cite{Hansen:2021fxi}. For a single scalar coupled to conformal gravity  this program has been performed  at  leading and sub-leading order \cite{Baiguera:2022lsw}. On the other hand, in the relativistic case the conformal technique has turned out to be very efficient in constructing general  matter couplings both with and without supersymmetry (see, e.g., \cite{Freedman:2012zz}). It is the aim of this work to apply the same conformal technique directly to the Carroll case. There are, however,  important differences with the relativistic case. One difference is that although there is a single conformal gravity theory, there are two types of massless Carroll scalars and correspondingly  two types of Carroll gravity theories. Another important difference is that in the Carroll case there exists a special type of torsion tensor, called {\sl intrinsic torsion}, that is independent of any spin-connection. While in general relativity non-zero torsion usually occurs only after coupling to matter, in the Carroll case these intrinsic torsion tensors can occur even in the absence of matter. Setting them to zero would lead to constraints on the geometry and therefore, they  cannot be ignored. In this work we will carefully follow these intrinsic torsion tensors and in particular show how they enter the transformation rules.

The idea behind the conformal technique is based upon the observation that there is a relation between dynamical matter and gravity. Restricting to a single scalar the relation in the relativistic case is as follows.  Coupling a dynamical scalar to conformal gravity and gauge-fixing dilatations by setting the scalar equal to one, one obtains the Einstein-Hilbert Lagrangian of general relativity. It also works the other way around. Replacing the Vierbein field in the Einstein-Hilbert Lagrangian by the product of a compensating scalar and a conformal Vierbein field and restricting to a flat spacetime one re-obtains the Lagrangian of a dynamical scalar. Restricting  to one scalar the conformal program leads to gravity without any additional matter coupling. In this sense this is only a first step in the conformal program. Once this step is completed one may obtain non-trivial matter couplings by replacing the single scalar by a function of $N$ scalars such that after gauge-fixing one is left with $N-1$ scalars coupled to gravity. In this work we will focus on the first step only.

Not every gravity theory gives rise within the conformal program to dynamical matter. For instance, in the relativistic case the Weyl tensor squared is already invariant under local dilatations and therefore cannot produce dynamical matter. We will find another example  of this in the Carroll case. Remarkably, we find that the conformal program also sometimes does not work the other way around: it is not always possible to couple dynamical matter to any   non-Lorentzian conformal gravity theory following  the structure of the underlying non-Lorentzian conformal algebra. We will see that  it works for conformal Carroll gravity but that it does not work in the usual way for conformal Galilei gravity. We note that even in the Carroll case the coupling is non-trivial and  only works due to some `lucky' coincidences.

Applying the conformal program to the Carroll case, we will find in this  work that 
\vskip .3truecm

\noindent 1. \ \ Coupling an  electric Carroll scalar to conformal Carroll gravity leads, upon gauge-fixing, to a non-conformal version of electric Carroll gravity.
\vskip .2truecm

\noindent  2. \ \ Coupling a magnetic  Carroll scalar to conformal Carroll gravity leads, upon gauge-fixing, to magnetic  Carroll gravity.
\vskip .2truecm

It is a bit ironic that the conformal program relates the more standard electric Carroll scalar to the more exotic electric Carroll gravity\,\footnote{For instance, electric Carroll gravity is built from intrinsic torsion tensors and therefore does not contain spin-connection fields.} whereas it relates the more exotic magnetic Carroll scalar\,\footnote{The Lagrangian of a magnetic Carroll scalar contains an additional Lagrange multiplier field. As we will see, this field disappears from the Lagrangian  after coupling to conformal Carroll gravity and gauge-fixing since it gets absorbed by a specific component of the independent Carroll boost spin-connection field.} to the more standard magnetic Carroll gravity theory\,\footnote{The magnetic Carroll gravity theory contains a term that is similar to the Einstein-Hilbert  term involving spin-connections.}.

This work is organized as follows. For the convenience of the reader we first review in section 2 the conformal technique in the relativistic case. To prepare the reader for the Carroll case, we use this relativistic case  to show how general torsion (in this case not intrinsic torsion) can arise in the transformation rules once it is included.  Furthermore, we emphasize that there are two ways of writing down the Lagrangian for dynamical matter that are equivalent in the sense that they differ by a total derivative but that, nevertheless, one of them  leads to simplifications in the calculations. Using the insights obtained from the relativistic case,
we perform in section 3 the conformal program to a single Carroll scalar both in the electric and magnetic case. We  discuss the corresponding Carroll gravity theories including the electric Carroll gravity theory that does not give rise to dynamical matter.
Next, in section 4 we discuss the issues that arise if one wishes to apply the conformal program to the Galilei case. Along the way, we will introduce a new Galilei gravity theory that contains an additional Lagrange multiplier field. Finally, in the Conclusions we discuss a few extensions of the present work.

\section{A Conformal Approach to General Relativity} \label{sec:relsec}

The purpose of this section is to explain the techniques that we wish to apply to Carroll gravity, in the familiar context of general relativity. In particular, we will show how the Einstein-Hilbert (EH) action can be obtained from that of a conformally coupled compensating scalar (i.e., a massless scalar coupled to suitable gauge fields of the conformal algebra), by gauge-fixing those conformal symmetries that are not part of the Poincar\'e algebra. These superfluous conformal symmetries are the dilatations, as well as the special conformal transformations whose role we will clarify. The scalar field is called {\sl compensating}, since its conformally coupled action can be re-obtained from the EH action by substituting the Vielbein in the latter with an appropriate product of a conformal Vielbein and the compensating scalar. The scalar thus compensates for the non-invariance of the conformal Vielbein under dilatations so that their product produces the scale-invariant Vielbein of general relativity.

Before discussing this compensating mechanism we first discuss in subsection \ref{2.1} the gauging of the conformal algebra\, \footnote{For early literature on gauging space-time symmetry algebras, see e.g., \cite{Chamseddine:1976bf,Kaku:1977pa}} and in subsection \ref{2.2} how to obtain a minimal representation of conformal algebra gauge fields. In the final subsection \ref{2.3} we will use this information to show how general relativity emerges from a conformally coupled compensating scalar.

\subsection{Gauging the conformal algebra} \label{2.1}

 The relativistic conformal algebra is spanned by the generators of translations $P_{\hat{A}}$, Lorentz transformations $M_{\hat{A}\hat{B}} =  - M_{\hat{B}\hat{A}}$, special conformal transformations $K_{\hat{A}}$ and dilatations $D$. Here, the indices $\hat{A}$, $\hat{B}$ take on the values $0, 1, \cdots, D-1$.\,\footnote{In this section, we use a hatted index $\hat{A}$, since the unhatted index $A$ will be used as a spatial index in the Carrollian case.} The non-zero commutation relations of the conformal algebra are given by 
\begin{alignat}{2} \label{eq:relconfalg}
  \comm{M_{\hat{A}\hat{B}}}{M_{\hat{C}\hat{D}}} &= 4\, \eta_{[\hat{A}[\hat{C}} M_{\hat{D}]\hat{B}]} \,, \qquad \qquad & \comm{P_{\hat{A}}}{M_{\hat{B}\hat{C}}} &= 2 \, \eta_{\hat{A}[\hat{B}} P_{\hat{C}]} \,, \nonumber \\
  \comm{K_{\hat{A}}}{M_{\hat{B}\hat{C}}} &= 2 \, \eta_{\hat{A}[\hat{B}} K_{\hat{C}]} \,, \qquad \qquad & \comm{P_{\hat{A}}}{K_{\hat{B}}} &= 2\, \eta_{\hat{A}\hat{B}} D + 2 \, M_{\hat{A}\hat{B}} \,, \nonumber \\
  \comm{D}{P_{\hat{A}}} &= P_{\hat{A}} \,, \qquad \qquad & \comm{D}{K_{\hat{A}}} &= - K_{\hat{A}} \,,
\end{alignat}
where $\eta_{\hat{A}\hat{B}}$ is the Minkowski metric for which we adopt the mostly plus signature convention. Introducing gauge fields $E_\mu{}^{\hat{A}}$ (called the Vielbein), $\Omega_\mu{}^{\hat{A}\hat{B}}$ (called the spin-connection), $F_\mu{}^{\hat{A}}$ and $B_\mu$, associated to respectively the generators $P_{\hat{A}}$, $M_{\hat{A}\hat{B}}$, $K_{\hat{A}}$ and $D$, the above commutation relations give rise to the following infinitesimal gauge transformation rules:
\begin{align}
  \label{eq:relconftrafos}
  \delta_0 E_{\mu}{}^{\hat{A}} &= \partial_\mu \zeta^{\hat{A}} - \zeta^{\hat{B}}\, \Omega_{\mu \hat{B}}{}^{\hat{A}} + \zeta^{\hat{A}}\, B_\mu  - \Lambda^{\hat{A}}{}_{\hat{B}} \, E_{\mu}{}^{\hat{B}} - \Lambda_D \, E_\mu{}^{\hat{A}} \,, \nonumber \\
  \delta_0 \Omega_{\mu}{}^{\hat{A}\hat{B}} &= \partial_\mu \Lambda^{\hat{A}\hat{B}} - 2 \, \Lambda^{[\hat{A}|}{}_{\hat{C}} \, \Omega_\mu{}^{\hat{C}|\hat{B}]} - 4 \, \Lambda_K{}^{[\hat{A}} \, E_{\mu}{}^{\hat{B}]}  - 4\, \zeta^{[\hat{A}}\, F_\mu{}^{\hat{B}]} \,, \nonumber \\
  \delta_0 F_{\mu}{}^{\hat{A}} &= \partial_\mu \Lambda_K{}^{\hat{A}} - \Lambda_{K \hat{B}}\, \Omega_\mu{}^{\hat{B} \hat{A}} - \Lambda_K{}^{\hat{A}} \, B_\mu - \Lambda^{\hat{A}}{}_{\hat{B}} \, F_{\mu}{}^{\hat{B}} + \Lambda_D \, F_\mu{}^{\hat{A}} \,, \nonumber \\
  \delta_0 B_\mu &= \partial_\mu \Lambda_D + 2\, \Lambda_K{}^{\hat{A}} \, E_{\mu \hat{A}} - 2\, \zeta^{\hat{A}}\, F_{\mu \hat{A}} \,.
\end{align}
Here, $\zeta^{\hat{A}}$, $\Lambda_D$, $\Lambda^{\hat{A}\hat{B}}$ and $\Lambda_K{}^{\hat{A}}$ are the parameters of translations, dilatations, Lorentz and special conformal transformations, respectively. In what follows, we will often refer to dilatations, Lorentz and special conformal transformations as \emph{homogeneous conformal transformations}. The action of the infinitesimal homogeneous conformal transformations of \eqref{eq:relconftrafos} will be denoted by the symbol $\delta_{\mathrm{hom},\, 0}$. We use the subscript ``0'' on $\delta_0$ and $\delta_{\mathrm{hom},\, 0}$ to indicate that these are not the final conformal transformations that are used to reconstruct the EH gravity action. Instead, the transformation rules used in the conformal approach are a modification of \eqref{eq:relconftrafos} with extra terms, as will be explained in the next section \ref{2.2}.

The curvatures that are defined in the following way
\begin{align}
  \label{eq:relconfcurvs}
  R_{\mu\nu}(P^{\hat{A}}) &\equiv 2\, \partial_{[\mu} E_{\nu]}{}^{\hat{A}} + 2\, \Omega_{[\mu}{}^{\hat{A}\hat{B}} \, E_{\nu] \hat{B}} + 2\, B_{[\mu} \, E_{\nu]}{}^{\hat{A}} \,, \nonumber \\
  R_{\mu\nu}(M^{\hat{A}\hat{B}}) &\equiv 2 \, \partial_{[\mu} \Omega_{\nu]}{}^{\hat{A}\hat{B}} + 2 \, \Omega_{[\mu|}{}^{[\hat{A}|}{}_{\hat{C}} \, \Omega_{|\nu]}{}^{\hat{C}|\hat{B}]} + 8 \, F_{[\mu}{}^{[\hat{A}} \, E_{\nu]}{}^{\hat{B}]} \,, \nonumber \\
  R_{\mu\nu}(K^{\hat{A}}) &\equiv 2 \partial_{[\mu} F_{\nu]}{}^{\hat{A}} + 2 \, \Omega_{[\mu}{}^{\hat{A} \hat{B}} \, F_{\nu] \hat{B}} - 2 \, B_{[\mu}\, F_{\nu]}{}^{\hat{A}} \,, \nonumber \\
  R_{\mu\nu}(D) &\equiv 2\, \partial_{[\mu} B_{\nu]} - 4\, F_{[\mu}{}^{\hat{A}} \, E_{\nu]\hat{A}} \,,
\end{align}
have the property that they transform into each other covariantly (i.e.~without the derivative of a parameter) under \eqref{eq:relconftrafos}. Explicitly, their transformations under \eqref{eq:relconftrafos} are given by:
\begin{align}
  \label{eq:curvtrafos}
  \delta_0 R_{\mu\nu}(P^{\hat{A}}) &=  - \zeta^{\hat{B}} \, R_{\mu\nu}(M_{\hat{B}}{}^{\hat{A}}) + \zeta^{\hat{A}} \, R_{\mu\nu}(D) - \Lambda^{\hat{A}}{}_{\hat{B}} \, R_{\mu\nu}(P^{\hat{B}}) - \Lambda_D \, R_{\mu\nu}(P^{\hat{A}}) \,, \nonumber \\
  \delta_0 R_{\mu\nu}(M^{\hat{A}\hat{B}}) &=  - 2 \, \Lambda^{[\hat{A}|}{}_{\hat{C}} \, R_{\mu\nu}(M^{\hat{C}|\hat{B}]}) - 4 \, \Lambda_K{}^{[\hat{A}} \, R_{\mu\nu}(P^{\hat{B}]}) - 4 \, \zeta^{[\hat{A}} \, R_{\mu\nu}(K^{\hat{B}]})\,, \nonumber \\
  \delta_0 R_{\mu\nu}(K^{\hat{A}}) &=  - \Lambda_{K \hat{B}}\, R_{\mu\nu}(M^{\hat{B} \hat{A}}) - \Lambda_K{}^{\hat{A}} \, R_{\mu\nu}(D) - \Lambda^{\hat{A}}{}_{\hat{B}} \, R_{\mu\nu}(K^{\hat{B}}) + \Lambda_D \, R_{\mu\nu}(K^{\hat{A}}) \,, \nonumber \\
  \delta_0 R_{\mu\nu}(D) &=  2\, \Lambda_K{}^{\hat{A}} \, R_{\mu\nu}(P_{\hat{A}}) - 2 \, \zeta^{\hat{A}} \, R_{\mu\nu}(K_{\hat{A}}) \,.
\end{align}
The covariant curvatures \eqref{eq:relconfcurvs} are not independent in the sense that they satisfy the following Bianchi identities:
\begin{align}
  \label{Bianchi}
  & D_{[\mu} R_{\nu\rho]}(P^{\hat{A}}) - R_{[\mu\nu}(M^{\hat{A}\hat{B}}) E_{\rho]\hat{B}} - R_{[\mu\nu}(D) E_{\rho]}{}^{\hat{A}} = 0 \,, \nonumber \\
  & D_{[\mu} R_{\nu\rho]}(M^{\hat{A}\hat{B}}) - 4 R_{[\mu\nu}(K^{[\hat{A}}) E_{\rho]}{}^{\hat{B}]} - 4 R_{[\mu\nu}(P^{[\hat{A}}) F_{\rho]}{}^{\hat{B}]} = 0 \,, \nonumber \\
  & D_{[\mu} R_{\nu\rho]}(K^{\hat{A}}) - R_{[\mu\nu} (M^{\hat{A}\hat{B}}) F_{\rho] \hat{B}} + R_{[\mu\nu}(D) F_{\rho]}{}^{\hat{A}} = 0 \,, \nonumber \\
  & \partial_{[\mu} R_{\nu\rho]}(D) + 2 R_{[\mu\nu}(K^{\hat{A}}) E_{\rho]\hat{A}} - 2 R_{[\mu\nu}(P^{\hat{A}}) F_{\rho]\hat{A}} = 0 \,.
\end{align}
Here we have used derivatives $D_{\mu}$ that are covariant with respect to Lorentz transformations and dilatations but not with respect to special conformal transformations.

\subsection{Imposing Constraints}\label{2.2}

In order to retrieve EH gravity via the conformal approach, the gauge fields and curvatures of the previous subsection must be supplemented with an additional ingredient. The main reason for this is that the multiplet of gauge fields \eqref{eq:relconftrafos} is not minimal, in the sense that it realizes the conformal algebra with more independent gauge fields than are needed for the construction of gravitational theories. The extra ingredient that is required to reduce the number of independent gauge fields consists of imposing the following two constraints on the curvatures:
\begin{equation}
  \label{eq:relconvconstr}
  R_{\mu\nu} (P^{\hat{A}}) = T_{\mu\nu}{}^{\hat{A}}  \qquad \qquad \text{and} \qquad \qquad R_{\mu \hat{B}} (M^{\hat{A}\hat{B}}) = 0 \,,
\end{equation}
where $T_{\mu\nu}{}^{\hat{A}}$ is an arbitrary tensor that, as will be seen shortly, corresponds to torsion. Ordinarily, when discussing the relativistic conformal approach, one only considers the case where the torsion vanishes, since one has in mind recovering general relativity. Including torsion in relativistic gravity is usually done by coupling to (fermionic) matter. Unlike Lorentzian geometry however, in its non-Lorentzian analogue part of the torsion is intrinsic. This intrinsic torsion can be present in non-Lorentzian gravity theories, even in the absence of matter couplings. Moreover, setting it to zero amounts to imposing constraints on the geometry. Choosing zero torsion will therefore be too restrictive in extensions of the conformal approach to Carroll gravity. In order to anticipate the Carrollian case, we will thus momentarily keep the torsion arbitrary to illustrate its effects, in particular how it deforms the conformal transformation rules. We will switch focus to the usual relativistic zero torsion case in the next section, when discussing how EH gravity arises from the conformal approach.

The constraints \eqref{eq:relconvconstr} are often called ``conventional'', since they can be used to solve the spin-connection $\Omega_\mu{}^{\hat{A}\hat{B}}$ and the special conformal gauge field $F_{\mu}{}^{\hat{A}}$ in terms of the Vielbein $E_\mu{}^{\hat{A}}$, the dilatation gauge field $B_\mu$ and the torsion $T_{\mu\nu}{}^{\hat{A}}$ as follows:
\begin{align} \label{eq:OmFsols}
  \Omega_\mu{}^{\hat{A}\hat{B}} (E,B,T) &= \Omega_\mu{}^{\hat{A}\hat{B}} (E,T) + 2\, E_\mu{}^{[\hat{A}} E^{\hat{B}]\nu} B_\nu \,, \nonumber \\
  F_{\mu}{}^{\hat{A}}(E,B,T) &= - \frac{1}{2 (D-2)} \left[R^\prime_{\mu \hat{B}} (M^{\hat{A}\hat{B}}) - \frac{1}{2(D-1)}\, E_\mu{}^{\hat{A}}\, R^\prime_{\hat{B} \hat{C}} (M^{\hat{B} \hat{C}}) \right] \,,
\end{align}
where $\Omega_\mu{}^{\hat{A}\hat{B}} (E,T)$ is the usual torsionful spin-connection
\begin{align} \label{eq:Omtorsion}
  \Omega_\mu{}^{\hat{A}\hat{B}} (E,T) &= E^{[\hat{A}|\nu} \left( 2 \partial_{[\mu} E_{\nu]}{}^{|\hat{B}]} - T_{\mu\nu}{}^{|\hat{B}]} \right) - \frac12 E_{\mu \hat{C}}\, E^{\hat{A}\nu}\, E^{\hat{B}\rho} \left(2 \partial_{[\nu} E_{\rho]}{}^{\hat{C}} - T_{\nu\rho}{}^{\hat{C}} \right) \,,
\end{align}
and $R^\prime_{\mu\nu}(M^{\hat{A}\hat{B}})$ is the Lorentz transformation curvature with the term containing the special conformal gauge field deleted:
\begin{align}
  R^\prime_{\mu\nu}(M^{\hat{A}\hat{B}}) &= 2 \, \partial_{[\mu} \Omega_{\nu]}{}^{\hat{A}\hat{B}}(E,B,T) + 2 \, \Omega_{[\mu|}{}^{[\hat{A}|}{}_{\hat{C}}(E,B,T) \, \Omega_{|\nu]}{}^{\hat{C}|\hat{B}]}(E,B,T) \,.
\end{align}
In what follows, it will be understood that the spin-connection $\Omega_\mu{}^{\hat{A}\hat{B}}$ and special conformal gauge field $F_\mu{}^{\hat{A}}$ are given by the dependent expressions $\Omega_\mu{}^{\hat{A}\hat{B}}(E,B,T)$ and $F_\mu{}^{\hat{A}}(E,B,T)$ of \eqref{eq:OmFsols}, whenever they appear in covariant derivatives or curvatures.

The transformation rules \eqref{eq:relconftrafos} typically need to be modified after the conventional constraints \eqref{eq:relconvconstr} are imposed. Let us first consider the homogeneous conformal transformations, whose modified infinitesimal version we will denote by $\delta_{\mathrm{hom}}$. Their action on the independent fields $E_\mu{}^{\hat{A}}$ and $B_\mu$ is assumed to be the same as that given in \eqref{eq:relconftrafos}. The rules $\delta_{\mathrm{hom}} \Omega_\mu{}^{\hat{A}\hat{B}}(E,B,T)$ and $\delta_{\mathrm{hom}} F_\mu{}^{\hat{A}}(E,B,T)$ of the dependent fields then follow from varying their explicit expressions \eqref{eq:OmFsols}.\,\footnote{This requires that one also assigns a homogeneous conformal transformation rule $\delta_{\mathrm{hom}} T_{\mu\nu}{}^{\hat{A}}$ for the torsion tensor. In what follows, we will assume that $T_{\mu\nu}{}^{\rho}$ is invariant under homogeneous conformal transformations, so that $\delta_{\mathrm{hom}} T_{\mu\nu}{}^{\hat{A}} = -\Lambda^{\hat{A}}{}_{\hat{B}} T_{\mu\nu}{}^{\hat{B}} - \Lambda_D T_{\mu\nu}{}^{\hat{A}}$.} The result of this does in general not coincide with the homologous rules given in \eqref{eq:relconftrafos}. This happens in particular when the conventional constraints \eqref{eq:relconvconstr} that are used to solve for the dependent fields are not left invariant by \eqref{eq:relconftrafos}. The transformation rules \eqref{eq:relconftrafos} of the dependent fields then acquire extra terms, whose role is to ensure that the constraints \eqref{eq:relconvconstr} are invariant under the thus modified rules. 

In the present case, one finds that the homogeneous conformal transformations of the independent and dependent gauge fields are, after imposing \eqref{eq:relconvconstr}, given by:
\begin{align} \label{eq:homreltrafos}
  \delta_{\mathrm{hom}} E_\mu{}^{\hat{A}} &= - \Lambda^{\hat{A}}{}_{\hat{B}} \, E_{\mu}{}^{\hat{B}} - \Lambda_D \, E_\mu{}^{\hat{A}} \,, \nonumber \\
  \delta_{\mathrm{hom}} B_\mu &= \partial_\mu \Lambda_D + 2\, \Lambda_K{}^{\hat{A}} \, E_{\mu \hat{A}} \,, \nonumber \\                                          
  \delta_{\mathrm{hom}} \Omega_{\mu}{}^{\hat{A}\hat{B}}(E,B,T) &= \partial_\mu \Lambda^{\hat{A}\hat{B}} - 2 \, \Lambda^{[\hat{A}|}{}_{\hat{C}} \, \Omega_\mu{}^{\hat{C}|\hat{B}]}(E,B,T) - 4 \, \Lambda_K{}^{[\hat{A}} \, E_{\mu}{}^{\hat{B}]} \,, \nonumber \\
  \delta_{\mathrm{hom}} F_{\mu}{}^{\hat{A}}(E,B,T) &= \partial_\mu \Lambda_K{}^{\hat{A}} - \Lambda_{K \hat{B}}\, \Omega_\mu{}^{\hat{B} \hat{A}}(E,B,T) - \Lambda_K{}^{\hat{A}} \, B_\mu - \Lambda^{\hat{A}}{}_{\hat{B}} \, F_{\mu}{}^{\hat{B}}(E,B,T) \nonumber \\ & \quad + \Lambda_D \, F_\mu{}^{\hat{A}}(E,B,T) + \frac{2}{D-2} \Lambda_K{}^{\hat{C}} \left[ \delta_{\hat{C}}^{[\hat{A}} T_{\mu \hat{B}}{}^{\hat{B}]} - \frac{1}{2 (D-1)} E_\mu{}^{\hat{A}}  T_{\hat{C}\hat{B}}{}^{\hat{B}} \right] \,.
\end{align}
Note that the first constraint of \eqref{eq:relconvconstr} is invariant under \eqref{eq:relconftrafos} and as a result the transformation rule of the dependent spin-connection $\Omega_\mu{}^{\hat{A}\hat{B}}(E,B,T)$ agrees with that of \eqref{eq:relconftrafos}. The second constraint of \eqref{eq:relconvconstr} by contrast transforms non-trivially under special conformal transformations:
\begin{align}
  \delta_{\mathrm{hom},\, 0} R_{\mu \hat{B}}(M^{\hat{A}\hat{B}}) \approx -4 \Lambda_K{}^{[\hat{A}} T_{\mu \hat{B}}{}^{\hat{B}]} \,,
\end{align}
where $\approx$ has the meaning of ``equality up to the conventional constraints \eqref{eq:relconvconstr}''. Consequently, the transformation rule of $F_\mu{}^{\hat{A}}(E,B,T)$  picks up extra special conformal transformation terms with respect to \eqref{eq:relconftrafos}.

Let us now discuss how also the translation transformations of \eqref{eq:relconftrafos} are modified due to the conventional constraints \eqref{eq:relconvconstr}. In this regard, it should be noted that translations are usually not considered as an independent symmetry, but rather as a combination of homogeneous conformal transformations and diffeomorphisms. Indeed, an infinitesimal diffeomorphism on any gauge field of the conformal algebra can be decomposed into homogeneous conformal transformations plus extra terms. These extra terms correspond to the translation rules of \eqref{eq:relconftrafos}, supplemented with curvature- and torsion-dependent contributions and it is thus sensible to view them as modified translation symmetries. In particular, any action that is invariant under diffeomorphisms and homogeneous conformal transformations is also automatically invariant under these modified translations. Note that generically the translation rules \eqref{eq:relconftrafos} of both the dependent and the independent fields receive extra terms. This is unlike what happens for the homogeneous conformal transformations that are only altered for the dependent fields.

Let us illustrate the above discussion by considering the following rewriting of the action of an infinitesimal diffeomorphism $\delta_\xi$ with parameter $\xi^\mu$ on the Vielbein $E_\mu{}^{\hat{A}}$:
\begin{align}
  \delta_\xi E_\mu{}^{\hat{A}} &= \partial_\mu \xi^\nu E_\nu{}^{\hat{A}} + \xi^\nu \partial_\nu E_\mu{}^{\hat{A}} = \partial_\mu \big(\xi^\nu E_\nu{}^{\hat{A}}\big) + 2 \xi^\nu \partial_{[\nu} E_{\mu]}{}^{\hat{A}} \nonumber \\
  &= - \xi^\nu \Omega_\nu{}^{\hat{A}}{}_{\hat{B}}(E,B,T) E_\mu{}^{\hat{B}} - \xi^\nu B_\nu E_\mu{}^{\hat{A}}  + \partial_\mu \big(\xi^\nu E_\nu{}^{\hat{A}}\big) + \Omega_\mu{}^{\hat{A}\hat{B}}(E,B,T) \xi^\nu E_{\nu \hat{B}} + B_\mu \xi^\nu E_\nu{}^{\hat{A}} \nonumber \\ & \qquad + \xi^\nu E_\nu{}^{\hat{B}} T_{\hat{B}\mu}{}^{\hat{A}} \,,
\end{align}
where in the last step we used the first of the conventional constraints \eqref{eq:relconvconstr}. Comparing the last line with \eqref{eq:relconftrafos}, we see that the first two terms correspond to homogeneous conformal transformations (namely a Lorentz transformation and dilatation with field-dependent parameters $\xi^\mu \Omega_\mu{}^{\hat{A}\hat{B}}(E,B,T)$ and $\xi^\mu B_\mu$ respectively) acting on $E_\mu{}^{\hat{A}}$. The remaining terms can then be identified with the action \eqref{eq:relconftrafos} of a translation on $E_\mu{}^{\hat{A}}$ with field-dependent parameter $\xi^\mu E_\mu{}^{\hat{A}}$, supplemented with an extra torsion contribution. This thus motivates defining the translation rule of $E_\mu{}^{\hat{A}}$ in the presence of the conventional constraints \eqref{eq:relconvconstr} as
\begin{align} \label{eq:modtrafoE}
\delta_\zeta E_\mu{}^{\hat{A}} &= \partial_\mu \zeta^{\hat{A}} + \Omega_\mu{}^{\hat{A}\hat{B}}(E,B,T) \zeta_{\hat{B}} + B_\mu \zeta^{\hat{A}} + \zeta^{\hat{B}} T_{\hat{B}\mu}{}^{\hat{A}} \,.
\end{align}
More generally, a modified translation with parameter $\zeta^{\hat{A}}$ is defined as the result of subtracting the homogeneous conformal transformations \eqref{eq:homreltrafos} with parameters  $\Lambda^{\hat{A}\hat{B}} = \zeta^{\hat{C}} E_{\hat{C}}{}^\mu \Omega_\mu{}^{\hat{A}\hat{B}}(E,B,T)$, $\Lambda_D = \zeta^{\hat{A}} E_{\hat{A}}{}^\mu B_\mu$ and $\Lambda_K{}^{\hat{A}} = \zeta^{\hat{B}} E_{\hat{B}}{}^\mu F_\mu{}^{\hat{A}}(E,B,T)$, from an infinitesimal diffeomorphism with parameter $\zeta^{\hat{A}} E_{\hat{A}}{}^\mu$. For $B_\mu$, $\Omega_\mu{}^{\hat{A}\hat{B}}(E,B,T)$ and $F_\mu{}^{\hat{A}}(E,B,T)$, this gives:
\begin{align} \label{eq:modtrafoBOmF}
  \delta_\zeta B_\mu &= - 2 \zeta^{\hat{A}} F_{\mu \hat{A}}(E,B,T) + \zeta^{\hat{A}} R_{\hat{A}\mu}(D) \,, \nonumber \\
  \delta_\zeta \Omega_\mu{}^{\hat{A}\hat{B}}(E,B,T) &= - 4 \zeta^{[\hat{A}} F_\mu{}^{\hat{B}]}(E,B,T) + \zeta^{\hat{C}} R_{\hat{C}\mu}(M^{\hat{A}\hat{B}}) \,, \nonumber \\
  \delta_\zeta F_\mu{}^{\hat{A}}(E,B,T) &= \zeta^{\hat{B}} R_{\hat{B}\mu}(K^{\hat{A}}) - \frac{2}{D-2} \zeta^{\hat{C}} F_{\hat{C}}{}^{[\hat{A}}(E,B,T) T_{\mu \hat{B}}{}^{\hat{B}]} \nonumber \\ & \qquad + \frac{1}{(D-1)(D-2)} E_\mu{}^{\hat{A}} \zeta^{\hat{D}} F_{\hat{D}}{}^{[\hat{B}}(E,B,T) T_{\hat{B}\hat{C}}{}^{\hat{C}]}  \,.
\end{align}
By making use of the Bianchi identities \eqref{Bianchi}, one can show explicitly that these translation rules of the dependent fields $\Omega_\mu{}^{\hat{A}\hat{B}}(E,B,T)$ and $F_\mu{}^{\hat{A}}(E,B,T)$ agree with those obtained from varying their expressions \eqref{eq:OmFsols} under \eqref{eq:modtrafoE} and the first rule of \eqref{eq:modtrafoBOmF}. In checking this, one assumes that a translation of the torsion tensor $T_{\mu\nu}{}^{\hat{A}}$ is also given by the action of a diffeomorphism minus that of a Lorentz transformation and dilatation with the above mentioned field-dependent parameters:
\begin{align}
  \label{eq:translT}
  \delta_\zeta T_{\mu\nu}{}^{\hat{A}} &= 2 \partial_{[\mu|}\left(\zeta^{\hat{B}} E_{\hat{B}}{}^\rho \right) T_{\rho|\nu]}{}^{\hat{A}} + \zeta^{\hat{B}} E_{\hat{B}}{}^{\rho} \partial_\rho T_{\mu\nu}{}^{\hat{A}} + \zeta^{\hat{C}} E_{\hat{C}}{}^\rho \Omega_\rho{}^{\hat{A}}{}_{\hat{B}}(E,B,T) T_{\mu\nu}{}^{\hat{B}} \nonumber \\ & \qquad + \zeta^{\hat{B}} E_{\hat{B}}{}^\rho B_\rho T_{\mu\nu}{}^{\hat{A}} \,. 
\end{align}

The independent gauge field $B_\mu$ plays a special role since it transforms with a shift under a special conformal transformation. In fact, it is the only field that transforms under a special conformal transformation. The Vielbein field $E_\mu{}^{\hat{A}}$ is invariant while the dependent gauge fields $\Omega_{\mu}{}^{\hat{A}\hat{B}}(E,B,T)$ and $F_{\mu}{}^{\hat{A}}(E,B,T)$ only transform under special conformal transformations due to their dependence on $B_\mu$. This means that $B_\mu$ is a Stueckelberg field that can be set to zero by fixing the special conformal transformations. The special conformal symmetry expresses the property that one can realize local dilatations without using a corresponding dilatation gauge field. The only independent gauge field left is therefore the Vielbein $E_\mu{}^{\hat{A}}$. As it stands this gauge field has $D^2$ field degrees of freedom. Due to the $D$ diffeomorphisms, the $\tfrac{1}{2}D(D-1)$ Lorentz rotations and the dilatation the number of independent degrees of freedom is given by $D^2 -D -\tfrac{1}{2}D(D-1) -1 = \tfrac{1}{2}(D+1)(D-2)$ which is precisely the number of helicity states of an irreducible massive spin-2 representation.

This counting of degrees of freedom can also be seen  by considering the independent curvature components. Substituting the conventional constraints \eqref{eq:relconvconstr} into the Bianchi identities \eqref{Bianchi} one finds after taking traces of the first and second Bianchi identity the following additional curvature constraints (for $D \neq 2$ and $D \neq 3$):
\begin{align} \label{eq:Bianchiconseq}
R_{\hat{A}\hat{B}}(D) &= \frac{1}{(D-2)} \left(3 D_{[\hat{C}} T_{\hat{A}\hat{B}]}{}^{\hat{C}} + T_{\hat{A}\hat{B}}{}^{\hat{C}} T_{\hat{C}\hat{D}}{}^{\hat{D}} \right)\,,  \qquad \qquad \qquad \qquad \text{and} \nonumber \\ R_{\hat{A}\hat{B}}(K^{\hat{C}}) &= \frac{1}{2 (D-3)} \left( D_{\hat{D}}R_{\hat{A}\hat{B}}(M^{\hat{C}\hat{D}}) - 2 T_{\hat{D}[\hat{A}}{}^{\hat{E}} R_{\hat{B}]\hat{E}}(M^{\hat{C}\hat{D}}) - 12 T_{[\hat{A}\hat{B}}{}^{[\hat{C}} F_{\hat{D]}}{}^{\hat{D}]} \right) \nonumber \\ & \quad \, + \frac{1}{2(D-2)(D-3)} \left(-T_{\hat{D}\hat{E}}{}^{\hat{F}} R_{\hat{F}[\hat{A}}(M^{\hat{D}\hat{E}}) + 4 T_{\hat{D}\hat{E}}{}^{[\hat{D}} F_{[\hat{A}}{}^{\hat{E}]} - 8 F_{\hat{D}}{}^{[\hat{D}} T_{\hat{E}[\hat{A}}{}^{\hat{E}]} \right) \delta_{\hat{B}]}^{\hat{C}} \,.
\end{align}
Therefore, the only independent curvature left is the Lorentz curvature. From the conventional constraints and the Bianchi identities, it follows that this Lorentz curvature satisfies a set of algebraic and differential identities (which we refrain from giving here) that reduce the $(\tfrac{1}{2}D(D-1))^2$ components to $\tfrac{1}{2}(D+1)(D-2)$ independent ones.

In the rest of this section \ref{sec:relsec}, we will set the torsion equal to zero to make contact with general relativity. The conventional constraints \eqref{eq:relconvconstr} are then invariant under \eqref{eq:relconftrafos}. As a consequence, the homogeneous conformal transformation rules of the dependent spin-connection and special conformal gauge field are not modified with respect to \eqref{eq:relconftrafos}, as is seen explicitly by setting $T_{\mu\nu}{}^{\hat{A}}$ equal to zero in \eqref{eq:homreltrafos}. From \eqref{eq:modtrafoE}, one sees that also the translation rule of the Vielbein coincides with that given in \eqref{eq:relconftrafos}. The same is true for the dilatation gauge field, since the curvature contribution in its translation rule \eqref{eq:modtrafoBOmF} vanishes as a consequence of Bianchi identities (see the first of \eqref{eq:Bianchiconseq} for $T_{\mu\nu}{}^{\hat{A}} = 0$). However, even in the absence of torsion, the translation rules of the spin-connection and special conformal gauge fields require modifications with curvature terms.

In what follows, we will denote the explicit solutions \eqref{eq:OmFsols} for the dependent spin-connection and special conformal gauge field in the zero torsion case by $\Omega_\mu{}^{\hat{A}\hat{B}}(E,B)$ and $F_\mu{}^{\hat{A}}(E,B)$. The Levi-Civita spin-connection, obtained by setting $T_{\mu\nu}{}^{\hat{A}} = 0$ in \eqref{eq:Omtorsion}, will be denoted by $\Omega_\mu{}^{\hat{A}\hat{B}}(E)$. For future reference, we note that, assuming zero torsion, we can relate $R^\prime_{\mu\nu}(M^{\hat{A}\hat{B}})$ to the usual Riemann tensor $R_{\mu\nu}{}^{\hat{A}\hat{B}}(E)$ in the following way, by writing out its $B_\mu$--dependence:
\begin{align} \label{eq:RpB}
  R^\prime_{\mu\nu}(M^{\hat{A}\hat{B}}) &= R_{\mu\nu}{}^{\hat{A}\hat{B}}(E) + 4\, E_{[\nu}{}^{\hat{A}}\, D_{\mu]}(E) B^{\hat{B}]} + 4\, E_{[\mu}{}^{[\hat{A}}\, B^{\hat{B}]} \, B_{\nu]} - 2 \, E_{[\mu}{}^{[\hat{A}}\, E_{\nu]}{}^{\hat{B}]} \, B^{\hat{C}} \, B_{\hat{C}} \,.
\end{align}
 Both the Riemann tensor $R_{\mu\nu}{}^{\hat{A}\hat{B}}(E)$ and the covariant derivative $D_\mu(E) B^{\hat{A}}$ are constructed with the usual Levi-Civita spin-connection $\Omega_\mu{}^{\hat{A}\hat{B}}(E)$:
\begin{align} \label{eq:RiemDE}
  R_{\mu\nu}{}^{\hat{A}\hat{B}}(E) &= 2 \, \partial_{[\mu} \Omega_{\nu]}{}^{\hat{A}\hat{B}}(E) + 2 \, \Omega_{[\mu|}{}^{[\hat{A}|}{}_{\hat{C}}(E) \, \Omega_{|\nu]}{}^{\hat{C}|\hat{B}]}(E) \,, \nonumber \\
  D_\mu(E) B^{\hat{A}} &= \partial_\mu B^{\hat{A}} + \Omega_\mu{}^{\hat{A}\hat{B}}(E) \, B_{\hat{B}} \,.
\end{align}
Note that the covariant derivative $D_\mu(E)$ is covariant with respect to the Lorentz transformations only, unlike the covariant derivative $D_\mu$ that we used in the Bianchi identities \eqref{Bianchi}.

\subsection{Einstein-Hilbert gravity from the compensating mechanism} \label{2.3}

We will now show how the conformal approach can be used to obtain the EH Lagrangian of general relativity
\begin{align} \label{eq:LEH}
  \mathcal{L}_{\mathrm{EH}} &= \frac{1}{2 \kappa^2} E R \,,
\end{align}
with $\kappa^2$ the gravitational coupling constant, in a two-step process. First, one couples a compensating scalar to the independent and dependent gauge fields of the conformal algebra that were discussed in the previous sections. The action of this scalar is determined by requiring invariance under the homogeneous conformal transformations and, manifestly, under diffeomorphisms. As discussed in the previous section, this then also guarantees invariance under translations, under the proviso that these act on the gauge fields with the modifications in \eqref{eq:modtrafoE} and \eqref{eq:modtrafoBOmF}. Einstein-Hilbert gravity is recovered next, by gauge-fixing the dilatations and special conformal transformations.

As starting point, we consider the Lagrangian of a free, massless scalar field $\phi$ in flat spacetime: 
\begin{align} \label{eq:Lscalar}
  \mathcal{L}_{\mathrm{scalar}} = -\frac12 \,  \phi \, \partial^{\hat A}\partial_{\hat A} \phi \,.
\end{align}
Note that the sign in front of this Lagrangian is the opposite of the usual, physical one. This will guarantee that the EH action comes out with the correct sign. We then assume that the transformation rule of the scalar field $\phi$ under local homogeneous conformal transformations is given by a dilatation with scaling weight $w$:
\begin{align}
  \label{eq:dilphirel}
  \delta_{\mathrm{hom}} \phi = w\, \Lambda_D \, \phi \,,
\end{align}
and we replace the ordinary derivatives in \eqref{eq:Lscalar} by covariant ones, by coupling to the gauge fields of the conformal algebra, discussed in the previous sections. The first order derivative of $\phi$ that transforms covariantly (i.e., without a derivative of a parameter) under homogeneous conformal transformations is defined by:
\begin{align}
  \label{eq:Dphi}
  D_{\hat{A}} \phi \equiv E_{\hat{A}}{}^\mu \left(\partial_\mu - w\, B_\mu \right) \phi \,.
\end{align}
Note that the second term in this covariant derivative is related to the transformation rule \eqref{eq:dilphirel} by replacing the dilatation parameter $\Lambda_D$ by (minus) the corresponding gauge field $B_\mu$. The covariant derivative $D_{\hat{A}} \phi$ transforms as follows under homogeneous conformal transformations:
\begin{align} \label{eq:Dphitrafo}
  \delta_{\mathrm{hom}} \left(D_{\hat{A}} \phi\right) = - \Lambda_{\hat{A}}{}^{\hat{B}} \, D_{\hat{B}} \phi + (w + 1)\, \Lambda_D \, D_{\hat{A}} \phi - 2\, w \, \Lambda_{K \hat{A}} \, \phi \,.
\end{align}
As a consequence, a covariant conformal box operator on $\phi$ can be introduced as:
\begin{align}
  \label{eq:cbox}
  \Box^C \phi \equiv E^{\hat{A}\mu} \left[\partial_\mu\left(D_{\hat{A}} \phi\right) + \Omega_{\mu \hat{A}}{}^{\hat{B}}(E,B) \, D_{\hat{B}} \phi - (w+1)\, B_\mu \, D_{\hat{A}} \phi + 2 \, w \, F_{\mu \hat{A}}(E,B) \, \phi \right] \,,
\end{align}
where the last three terms are obtained by replacing the gauge parameters on the right-hand-side of \eqref{eq:Dphitrafo} by (minus) their corresponding gauge fields. Explicit computation shows that the homogeneous conformal transformation rule of $\Box^C \phi$ is given by:
\begin{align} \label{eq:trafocbox}
  \delta_{\mathrm{hom}} \Box^C \phi = (w + 2) \, \Lambda_D \, \Box^C \phi + (2 D - 4 w - 4) \, \Lambda_K{}^{\hat{A}} \, D_{\hat{A}} \phi \,.
\end{align}

Replacing the flat space-time d'Alembertian in \eqref{eq:Lscalar} by $\Box^C$, one is led to the following Lagrangian of a conformally coupled scalar:
\begin{align} \label{eq:Lconf}
  \mathcal{L}_{\mathrm{conf}} = -\frac12 \, E\, \phi \, \Box^C \phi \,.
\end{align}
From the transformation rule \eqref{eq:trafocbox}, one sees that $\mathcal{L}_{\mathrm{conf}}$ is not invariant under special conformal transformations unless the weight $w$ is given by:
\begin{align}
  \label{eq:wval}
  w = \frac{D}{2} - 1 \,.
\end{align}
Since $E$, the determinant of the Vielbein $E_\mu{}^{\hat{A}}$, has scaling weight $-D$ under dilatations, this value of $w$ also renders $\mathcal{L}_{\mathrm{conf}}$ invariant under dilatations and thus under all homogeneous conformal symmetries.

The EH Lagrangian can be retrieved from $\mathcal{L}_{\mathrm{conf}}$ by gauge-fixing the special conformal transformations and dilatations. To fix the latter, we set $\phi$ equal to a constant $k$:
\begin{align} \label{eq:dilfix}
  \text{dilatation gauge-fixing :} \ \ \ \phi = k \,.
\end{align}
Imposing this gauge-fixing condition in $\mathcal{L}_{\mathrm{conf}}$, one obtains
\begin{align} \label{eq:DfixL}
  \mathcal{L}_{\mathrm{conf}} \ \xrightarrow{\phi = 1} \ \mathcal{L} &= \frac{w k^2}{2}\, E\, E^{\hat{A}\mu}\, \Big[ \left(\partial_\mu  B_{\hat{A}} + \Omega_{\mu \hat{A}}{}^{\hat{B}} (E,B)\, B_{\hat{B}}\right) - (w+1)\, B_\mu\, B_{\hat{A}} - 2 \, F_{\mu \hat{A}}(E,B)  \Big] \,, \nonumber \\
  &= \frac{w k^2}{2}\, E\, \Big[ D^{\hat{A}}(E) B_{\hat{A}} + (D - w - 2) B^{\hat{A}} B_{\hat{A}} - 2\, F_{\hat{A}}{}^{\hat{A}}(E,B)\Big] \,.
\end{align}
Since $B_{\hat{A}}$ transforms with a Stueckelberg shift $2 \Lambda_{K \hat{A}}$ under special conformal transformations, the field $B_{\hat{A}}$ can not appear in \eqref{eq:DfixL}. Indeed, using the explicit expression for $F_{\mu}{}^{\hat{A}}(E,B)$ in \eqref{eq:OmFsols} (with $T_{\mu\nu}{}^{\hat{A}} = 0$), along with \eqref{eq:RpB} and \eqref{eq:RiemDE}, one finds that:
\begin{align}
 - 2\, F_{\hat{A}}{}^{\hat{A}}(E,B) &= \frac{R}{2(D-1)} - D^{\hat{A}}(E) B_{\hat{A}} - \frac12 (D-2) B^{\hat{A}} \, B_{\hat{A}} \,,
\end{align}
where $R = R_{\hat{A}\hat{B}}{}^{\hat{A}\hat{B}}(E)$ is the Ricci scalar of general relativity. The Lagrangian \eqref{eq:DfixL} can thus be rewritten as
\begin{align}
  \mathcal{L} &= \frac{w k^2}{2} E\, \Big[\frac{R}{2 (D-1)} - \frac{1}{2}\left(2 w + 2 - D \right) B^{\hat{A}}\, B_{\hat{A}} \Big] \,.
\end{align}
The last term vanishes due to the parameter constraint \eqref{eq:wval} that guarantees conformal invariance of \eqref{eq:Lconf} and one finds that \eqref{eq:DfixL} is given by:
\begin{align} \label{eq:fullfixL}
  \mathcal{L} &=  \frac{w k^2}{4(D-1)} E\, R = \frac{k^2}{8} \frac{(D-2)}{(D-1)} E\, R \,.
\end{align}
Alternatively, one can immediately obtain \eqref{eq:fullfixL} from \eqref{eq:DfixL} by gauge-fixing the special conformal transformations via the gauge choice:
\begin{align} \label{eq:spconffix}
  \text{special conformal gauge-fixing :} \ \ \ B_{\hat{A}} = 0 \,.
\end{align}
Either way, one finds that, upon choosing the constant $k$ as:
\begin{align} \label{eq:kval}
  k = \frac{2}{\kappa} \sqrt{\frac{(D-1)}{(D-2)}} \,,
\end{align}
\eqref{eq:fullfixL} becomes the EH Lagrangian \eqref{eq:LEH}, showing that the latter can indeed be obtained as a partial gauge-fixing of the conformal matter Lagrangian \eqref{eq:Lconf}.

For later use, it is instructive to consider a slightly different starting point to derive the EH action from a compensating mechanism. Instead of starting from the Lagrangian \eqref{eq:Lscalar}, we could also start from the following Lagrangian in flat spacetime:
\begin{align} \label{eq:Lscalar2}
  \mathcal{L}^\prime_{\mathrm{scalar}} = \frac12 \,  \partial_{\hat A}\phi \partial^{\hat A} \phi \,,
\end{align}
that differs from \eqref{eq:Lscalar} by a total derivative. Following the same strategy of replacing derivatives by covariant ones we end up with the following Ansatz for the Lagrangian of a scalar coupled to gauge fields of the conformal algebra: 
\begin{align}
  \mathcal{L}^\prime_{\rm Ansatz} = \frac{E}{2} \, D_{\hat{A}} \phi\, D^{\hat{A}} \phi \,,
\end{align}
where $D_{\hat A}\phi$ has been defined in \eqref{eq:Dphi}. However, as it stands, this Lagrangian is not conformally invariant. In particular, under special conformal transformations one has
\begin{align}
  \delta_K \mathcal{L}^\prime_{\rm Ansatz} = E\, D^{\hat{A}} \phi \, \delta_K \left(D_{\hat{A}} \phi\right) = -2 w \, E\, \Lambda_{K \hat{A}} \, \phi \, D^{\hat{A}} \phi = - w\, E \, E_{\hat{A}}{}^{\mu} \, \Lambda_{K}{}^{\hat{A}} \, D_\mu(\phi^2) \,,
\end{align}
with $D_{\mu}(\phi^2) = \partial_\mu (\phi^2) - 2 \, w \, B_\mu \, \phi^2$ (since $\phi^2$ has dilatation weight $2 w$). To see how this variation can be cancelled, one partially integrates
\begin{align} \label{eq:partint}
  \delta_K \mathcal{L}^\prime_{\mathrm{Ansatz}} &= \, w \partial_\mu\left(E\, E_{\hat{A}}{}^{\mu}\right) \Lambda_K{}^{\hat{A}} \, \phi^2 + w \, E\, E_{\hat{A}}{}^{\mu}\, \partial_\mu \Lambda_K{}^{\hat{A}} \, \phi^2 + 2 \, w^2 \, E \, \Lambda_{K}{}^{\hat{A}} \, B_{\hat{A}}\, \phi^2 \nonumber \\
  &= w \, E\, E_{\hat{A}}{}^\mu \, \left(\delta_K F_\mu{}^{\hat{A}} (E,B)\right) \, \phi^2 + w \left(2 w + 2 -D\right) E\, \Lambda_K{}^{\hat{A}}\, B_{\hat{A}}\, \phi^2 \,,
\end{align}
where
\begin{align}
  \delta_K F_\mu{}^{\hat{A}} (E,B) &= \partial_\mu \Lambda_K{}^{\hat{A}} - \Lambda_{K \hat{B}}\, \Omega_\mu{}^{\hat{B} \hat{A}}(E,B) - \Lambda_K{}^{\hat{A}} \, B_\mu \,.
\end{align}
is an infinitesimal special conformal transformation of $F_\mu{}^{\hat{A}}(E,B)$ (see eq.~\eqref{eq:homreltrafos} with $T_{\mu\nu}{}^{\hat{A}} = 0$). In the first line of \eqref{eq:partint} we dropped a total derivative term, while in going to the second line we used the identity
\begin{align} \label{eq:partintrule}
  \partial_\mu \left(E\, E_{\hat{A}}{}^\mu \right)& = 2 \, E\, E_{\hat{A}}{}^{\mu} \, E_{\hat{B}}{}^\nu \, \partial_{[\mu} E_{\nu]}{}^{\hat{B}} \nonumber \\ &= -2 \, E\, E_{\hat{A}}{}^{\mu} \, E_{\hat{B}}{}^\nu \, \Omega_{[\mu}{}^{\hat{B}\hat{C}}(E,B) \, E_{\nu] \hat{C}} - 2 \, E \, E_{\hat{A}}{}^{\mu} \, E_{\hat{B}}{}^\nu \, B_{[\mu} \, E_{\nu]}{}^{\hat{B}} \,,
\end{align}
where the first line follows from $\partial_\mu E = E E_{\hat{A}}{}^\nu \partial_\mu E_{\nu}{}^{\hat{A}}$ and $\partial_\nu E_{\hat{A}}{}^\mu = - E_{\hat{A}}{}^\rho E_{\hat{B}}{}^\mu \partial_\nu E_{\rho}{}^{\hat{B}}$ and in the second line we have used the first of the conventional constraints \eqref{eq:relconvconstr}, with $T_{\mu\nu}{}^{\hat{A}} = 0$. Choosing the dilatation weight $w$ as in \eqref{eq:wval}, the last term of \eqref{eq:partint} vanishes. Since the remaining variation of $\mathcal{L}^\prime_{\mathrm{Ansatz}}$ can be cancelled by that of $- w\, E\, F_{\hat{A}}{}^{\hat{A}}(E,B) \, \phi^2$, adding this term to $\mathcal{L}^\prime_{\rm Ansatz}$ leads to the following  conformally invariant Lagrangian $\mathcal{L}^{\prime}_{\mathrm{conf}}$:
\begin{align}
  \mathcal{L}^{\prime}_{\mathrm{conf}} = \mathcal{L}^{\prime}_{\rm Ansatz} - w\, E\, F_{\hat{A}}{}^{\hat{A}}(E,B) \, \phi^2 = \frac{E}{2} \left[D_{\hat{A}} \phi D^{\hat{A}} \phi - 2 \, w\, F_{\hat{A}}{}^{\hat{A}}(E,B) \, \phi^2 \right] \,.
\end{align}
By partially integrating and using \eqref{eq:wval} and \eqref{eq:partintrule}, one can see that $\mathcal{L}^{\prime}_{\mathrm{conf}}$ is equal to the Lagrangian $\mathcal{L}_{\mathrm{conf}}$ given in \eqref{eq:Lconf}, up to a total derivative. The EH Lagrangian \eqref{eq:LEH} is then recovered from $\mathcal{L}^\prime_{\mathrm{conf}}$ by adopting the dilatation and special conformal gauge-fixing conditions of eqs.~\eqref{eq:dilfix} (with $k$ again given by \eqref{eq:kval}) and \eqref{eq:spconffix}.

This finishes our review of the conformal compensating technique in the relativistic case.

\section{A Conformal Approach to Carroll Gravity}

Here, we will adapt the conformal compensating technique discussed in the previous section to construct two different Carroll gravity theories. The existence of these two theories is related to the fact that there exist two massless Carroll scalar field theories that are called `electric' and `magnetic' in the literature \cite{Hansen:2021fxi,Henneaux:2021yzg,Perez:2021abf,Perez:2022jpr,Bergshoeff:2022eog}. In subsection \ref{ssec:confcarrgauging}, we will first discuss the gauging of the conformal Carroll algebra, which corresponds to an In\"on\"u-Wigner contraction of the relativistic conformal algebra. In the next subsection \ref{ssec:carrconstr}, we will derive a minimal multiplet of gauge fields of the conformal Carroll algebra, by solving suitable conventional curvature constraints to express some of the gauge fields in terms of the remaining independent ones. Finally, in the last subsection \ref{ssec:carrgravs}, we will construct conformally coupled versions of electric and magnetic Carroll scalars and adopt gauge-fixing conditions, giving rise to two different Carroll gravity theories.

\subsection{The Conformal Carroll Algebra and its Gauging} \label{ssec:confcarrgauging}

The commutation relations of the conformal Carroll algebra can be obtained by an In\"on\"u-Wigner contraction of the relativistic conformal algebra, given in \eqref{eq:relconfalg}. To perform this contraction we decompose the flat Lorentz index $\hat A$ as $\hat A = \{0,A\}$, where $0$ is a flat time index and $A = 1,\cdots, D-1$ are flat spatial indices. The relativistic conformal generators decompose correspondingly as
\begin{align}
 M_{\hat{A}\hat{B}} &\rightarrow M_{0A}, M_{AB} \,,\hskip 1.5truecm
 P_{\hat{A}} \rightarrow P_{0}, P_{A} \,,\hskip 1.5truecm
 K_{\hat{A}} \rightarrow K_{0}, K_{A} \,.
\end{align}
Next, we redefine the conformal generators with a contraction parameter $c$ as follows:
\begin{alignat}{3}\label{redefC}
  M_{0A} \ \ &= \ \ {c}^{-1}\, J_{0A} \,, \qquad \quad & M_{A B} \ \ &= \ \ J_{A B} \,, \qquad \quad & P_0 \ \ &= \ \ {c}^{-1} H \,, \qquad \quad P_A \ \ = \ \ P_A \,, \nonumber \\
  K_0 \ \ &= \ \ {c}^{-1}\, K \,, \qquad \quad &K_{A} \ \ &= \ \  \, K_{A} \,, \qquad \quad & D \ \ &= \ \ D \,.
\end{alignat}
The generators on the right-hand-side become those of the conformal Carroll algebra after taking the $c \to 0$ limit. In what follows, we will refer to the transformations associated to $J_{AB}$, $J_{0A}$, $K$/$K_A$ and $D$ as spatial rotations, Carroll boosts, singlet/vector special conformal transformations and dilatations, respectively. They constitute the subalgebra of homogeneous conformal Carroll transformations, whereas $H$ and $P_A$ correspond to time and spatial translations. Substituting the redefinitions \eqref{redefC}  into the relativistic commutation relations \eqref{eq:relconfalg} and taking the $c \to 0$ limit, we obtain the following non-zero commutation relations of the conformal Carroll algebra \cite{Nguyen:2023vfz,Afshar:2024llh}:
\begin{alignat}{3} \label{eq:CCA1}
\comm{J_{AB}}{J_{CD}} &= 4\, \delta_{[A[C} J_{D]B]} \,, \qquad & \comm{J_{AB}}{J_{0C}} &= 2\, \delta_{C[B|} J_{0| A]} \qquad & \comm{P_{A}}{J_{BC}} &= 2 \, \delta_{A[B} P_{C]} \,, \nonumber \\[.1truecm]
  \comm{P_{A}}{J_{0B}} &= - \, \delta_{AB}H \,, \qquad & \comm{P_{A}}{K_{B}} &= 2\, \left(\delta_{AB}D+J_{AB}\right) \,, \qquad & \comm{K_{A}}{J_{BC}} &= 2\,\delta_{A[B} K_{C]} \,, \nonumber \\[.1truecm]
\comm{K_{A}}{J_{0B}} &= - \, \delta_{AB}K \,, \qquad &  \comm{P_{A}}{K} &= -2 \, J_{0A} \,, \qquad & \comm{H}{K_{A}} &= 2\,J_{0A} \,, \nonumber \\[.1truecm]
 \comm{D}{P_{A}} &= P_{A} \,, \qquad & \comm{D}{H} &= H \,, \qquad & \comm{D}{K_{A}} &=- K_{A} \,, \nonumber \\[.1truecm]
\comm{D}{K} &= -K \,.
\end{alignat}

We then introduce the gauge fields $\tau_\mu$, $e_\mu{}^{A}$, $\omega_{\mu}{}^{AB}$, $\omega_{\mu}{}^{0A}$, $f_\mu$, $g_{\mu}{}^{A}$ and $b_{\mu}$, associated to the generators $H$, $P_{A}$, $J_{AB}$, $J_{0A}$, $K$, $K_{A}$ and $D$, respectively. From the algebra \eqref{eq:CCA1}, we find that these gauge fields transform under time/spatial translations, spatial rotations, Carroll boosts, singlet/vector special conformal transformations and dilatations, with respective gauge parameters $\zeta$/$\zeta^A$, $\lambda^{AB},\lambda^{0A},\lambda_K$/$\lambda_K{}^{A}$ and $\lambda_D$, as follows:
\begin{align}
  \label{eq:confcarrtrafos}
  \delta_0 \tau_\mu &= \tilde{D}_\mu \zeta - \lambda^{0A} e_{\mu A} - \lambda_D \tau_\mu \,, \nonumber \\
  \delta_0 e_\mu{}^A &= \tilde{D}_\mu \zeta^A - \lambda^{A}{}_{B} e_\mu{}^{B} - \lambda_D e_\mu{}^A \,, \nonumber \\
  \delta_0 \omega_\mu{}^{AB} &= \partial_\mu \lambda^{AB} - 2 \lambda^{[A|}{}_C \omega_\mu{}^{C|B]} - 4 \lambda_K{}^{[A} e_\mu{}^{B]}  -  4 \zeta^{[A} g_\mu{}^{B]}  \,, \nonumber \\
  \delta_0 \omega_\mu{}^{0A} &= \partial_\mu \lambda^{0A} + \omega_\mu{}^{A}{}_{B} \lambda^{0B} - \lambda^{A}{}_{B} \omega_\mu{}^{0B} + 2 \lambda_K{}^{A} \tau_\mu - 2 \lambda_K e_\mu{}^{A}  + \ 2  \zeta^A f_\mu - 2 \zeta g_\mu{}^{A} \,, \nonumber \\
  \delta_0 f_\mu &= \partial_\mu \lambda_K - \lambda_K b_\mu - \lambda^{0A} g_{\mu A} + \lambda_{K A} \omega_\mu{}^{0A} + \lambda_D f_\mu \,, \nonumber \\
  \delta_0 g_\mu{}^A &= \partial_\mu \lambda_K{}^A + \omega_\mu{}^{A}{}_{B} \lambda_K{}^B - \lambda_K{}^{A} b_\mu - \lambda^{A}{}_{B} g_\mu{}^{B} + \lambda_D g_\mu{}^{A} \,, \nonumber \\
  \delta_0 b_\mu &= \partial_\mu \lambda_D + 2 \lambda_K{}^{A} e_{\mu A}  -  2 \zeta^{A} g_{\mu A}  \,,
\end{align}
where
\begin{align} \label{eq:deftildeD}
  \tilde{D}_\mu \zeta \equiv \partial_\mu \zeta + \omega_\mu{}^{0A} \zeta_A + \zeta b_\mu \,, \qquad \qquad \tilde{D}_\mu \zeta^A \equiv \partial_\mu \zeta^A + \omega_\mu{}^{AB} \zeta_B + \zeta^A b_\mu \,.
\end{align}
We use a subscript ``0'' on $\delta_0$ to indicate that these are not the final conformal Carroll transformation rules that we will use to construct Carroll gravity theories. Like in the relativistic case, we will need to impose conventional constraints that will supplement these rules with extra terms and it is the thus modified transformations that will be employed in the compensating mechanism. These changes to \eqref{eq:confcarrtrafos} will be discussed in the next section \ref{ssec:carrconstr}. We will furthermore also make use of the symbol $\delta_{\mathrm{hom},\, 0}$ to specify the action of the homogeneous conformal Carroll transformations of \eqref{eq:confcarrtrafos}.

Starting from the conformal Carroll algebra \eqref{eq:CCA1}, one defines the following curvatures:
\begin{subequations} \label{eq:carrconfcurvs}
\begin{align}   
  R_{\mu\nu}(H) &= 2\, \partial_{[\mu} \tau_{\nu]}{} + 2\, \omega_{[\mu}{}^{0A} \, e_{\nu] A} + 2\, b_{[\mu} \, \tau_{\nu]} \,,  \\
  R_{\mu\nu}(P^{A}) &= 2\, \partial_{[\mu} e_{\nu]}{}^{A} + 2\, \omega_{[\mu}{}^{AB} \, e_{\nu] B} + 2\, b_{[\mu} \, e_{\nu]}{}^{A} \,,  \\
  R_{\mu\nu}(J^{AB}) &= 2 \, \partial_{[\mu} \omega_{\nu]}{}^{AB} + 2 \, \omega_{[\mu|}{}^{[A|}{}_{C} \, \omega_{|\nu]}{}^{C|B]} + 8 \, g_{[\mu}{}^{[A} \, e_{\nu]}{}^{B]} \,, \label{eq:RJAB} \\
  R_{\mu\nu}(J^{0A}) &= 2 \, \partial_{[\mu} \omega_{\nu]}{}^{0A} + 2 \, \omega_{[\mu|}{}^{A}{}_{B} \, \omega_{|\nu]}{}^{0 B} +4 \, f_{[\mu} \, e_{\nu]}{}^{A} -4g_{[\mu}{}^{A} \, \tau_{\nu]}\,, \label{eq:RJ0A} \\
  R_{\mu\nu}(K) &= 2 \partial_{[\mu} f_{\nu]} + 2 \, \omega_{[\mu}{}^{0A} \, g_{\nu]A} - 2 \, b_{[\mu}\, f_{\nu]} \,,  \\  
  R_{\mu\nu}(K^{A}) &= 2 \partial_{[\mu} g_{\nu]}{}^{A} + 2 \, \omega_{[\mu}{}^{AB} \, g_{\nu] B} - 2 \, b_{[\mu}\, g_{\nu]}{}^{A} \,,  \\
 R_{\mu\nu}(D) &= 2\, \partial_{[\mu} b_{\nu]} - 4\, g_{[\mu}{}^{A} \, e_{\nu]A} \,.
\end{align}
\end{subequations}
These curvatures transform covariantly under \eqref{eq:confcarrtrafos}; in particular, one has:
\begin{align}
   \label{eq:confcarrtranscurv}
  \delta_0 R_{\mu\nu}(H) &=  \zeta_A R_{\mu\nu}(J^{0A}) + \zeta R_{\mu\nu}(D) - \lambda^{0A} R_{\mu\nu}(P_A) - \lambda_D R_{\mu\nu}(H) \,, \nonumber \\
  \delta_0 R_{\mu\nu}(P^A) &= - \zeta_B R_{\mu\nu}(J^{BA}) + \zeta^A R_{\mu\nu}(D) - \lambda^{A}{}_{B} R_{\mu\nu}(P^B) - \lambda_D R_{\mu\nu}(P^A) \,, \nonumber \\
  \delta_0 R_{\mu\nu}(J^{AB}) &= - 2 \lambda^{[A|}{}_{C} R_{\mu\nu}(J^{C|B]}) - 4 \lambda_K{}^{[A} R_{\mu\nu}(P^{B]}) - 4 \zeta^{[A} R_{\mu\nu}(K^{B]}) \,, \nonumber \\
  \delta_0 R_{\mu\nu}(J^{0A}) &= - \lambda^{0B} R_{\mu\nu}(J_{B}{}^{A}) - \lambda^{A}{}_{B} R_{\mu\nu}(J^{0B}) + 2 \lambda_K{}^{A} R_{\mu\nu}(H) - 2 \lambda_K R_{\mu\nu}(P^A) \nonumber \\ & \qquad  + 2 \zeta^{A} R_{\mu\nu}(K) - 2 \zeta R_{\mu\nu}(K^{A}) \,, \nonumber \\
  \delta_0 R_{\mu\nu}(K) &= - \lambda_K R_{\mu\nu}(D) - \lambda^{0A} R_{\mu\nu}(K_A) + \lambda_{K A} R_{\mu\nu}(J^{0A}) + \lambda_D R_{\mu\nu}(K) \,, \nonumber \\
  \delta_0 R_{\mu\nu}(K^A) &= - \lambda_K{}^{B} R_{\mu\nu}(J_{B}{}^{A}) - \lambda_K{}^{A} R_{\mu\nu}(D) - \lambda^{A}{}_{B} R_{\mu\nu}(K^{B}) + \lambda_D R_{\mu\nu}(K^{A}) \,, \nonumber \\
  \delta_0 R_{\mu\nu}(D) &= 2 \lambda_K{}^{A} R_{\mu\nu}(P_{A}) - 2 \zeta^{A} R_{\mu\nu}(K_{A}) \,.
\end{align}
Furthermore, they satisfy the following Bianchi identities: 
\begin{align}
& D_{[\mu} R_{\nu\rho]}(H)+ R_{[\mu\nu}(P^{A})\omega_{\rho]}{}^{0}{}_{A}- R_{[\mu\nu}(J^{0A}) e_{\rho]A} - R_{[\mu\nu}(D) \tau_{\rho]} = 0 \,, \nonumber \\
    & D_{[\mu} R_{\nu\rho]}(P^{A}) - R_{[\mu\nu}(J^{AB}) e_{\rho]B} - R_{[\mu\nu}(D) e_{\rho]}{}^{A} = 0 \,, \nonumber \\
  & D_{[\mu} R_{\nu\rho]}(J^{AB}) - 4 R_{[\mu\nu}(K^{[A}) e_{\rho]}{}^{B]} - 4 R_{[\mu\nu}(P^{[A}) g_{\rho]}{}^{B]} = 0 \,, \nonumber \\
& D_{[\mu} R_{\nu\rho]}(J^{0A})-R_{[\mu\nu}(J^{AB})\omega_{\rho]}{}^{0}{}_{B}+2 R_{[\mu\nu}(K^{A}) \tau_{\rho]} - 2 R_{[\mu\nu}(K) e_{\rho]}{}^{A} \nonumber\\
&\qquad +2R_{[\mu\nu}(P^A)f_{\rho]}- 2 R_{[\mu\nu}(H) g_{\rho]}{}^{A} = 0 \,, \nonumber \\
   & D_{[\mu} R_{\nu\rho]}(K) +R_{[\mu\nu}(K^{A})\omega_{\rho]}{}^{0}{}_{A}- R_{[\mu\nu} (J^{0A}) g_{\rho] A} + R_{[\mu\nu}(D) f_{\rho]}{} = 0 \,, \nonumber \\  
     & D_{[\mu} R_{\nu\rho]}(K^{A}) - R_{[\mu\nu} (J^{AB}) g_{\rho] B} + R_{[\mu\nu}(D) g_{\rho]}{}^{A} = 0 \,, \nonumber \\
  & \partial_{[\mu} R_{\nu\rho]}(D) + 2 R_{[\mu\nu}(K^{A}) e_{\rho]A} - 2 R_{[\mu\nu}(P^{A}) g_{\rho]A} = 0 \,.
\end{align}
where the derivatives $D_\mu$ are covariant with respect to spatial rotations and dilatations but not with respect to Carroll boosts and singlet/vector special conformal transformations since the latter do not act in a homogeneous manner.

We will often need the dual Vielbeine $(\tau^{\mu},e_{A}{}^{\mu})$ that are defined by the following duality relations:
\begin{align} \label{invVielbdef}
    \tau^{\mu}\tau_\mu &= 1\,, & \tau^{\mu}e_{\mu}{}^{A} &= 0\,, & \tau_{\mu}e_A{}^{\mu} &= 0\,,\notag \\[.1truecm]
e_{\mu}{}^{A}e_{B}{}^{\mu} &= \delta^{A}_B\,, & e_{\mu}{}^{A}e_{A}{}^{\nu} &= \delta^{\mu}_{\nu}-\tau^{\mu}\tau_\nu\,.
\end{align}
These dual Vielbeine transform as follows under \eqref{eq:confcarrtrafos}:
\begin{align}\label{inverseV}
    \delta_0 \tau^{\mu} &= -\tau^\mu \tilde{D}_0 \zeta - e_A{}^\mu \tilde{D}_0 \zeta^A + \lambda_D \tau^\mu \,, \nonumber \\ \delta_0 e_{A}{}^{\mu} &= -\tau^\mu \tilde{D}_A \zeta - e_B{}^\mu \tilde{D}_A \zeta^B - \lambda_A{}^{B}e_B{}^{\mu} + \lambda^{0}{}_{A}\tau^{\mu} + \lambda_D e_{A}{}^{\mu} \,.
\end{align}
In the following, we will use $\tau^\mu$ and $e_A{}^\mu$ to convert a curved index $\mu$ into flat indices $0$ and $A$, respectively.
For instance, given a one-form $X_\mu$ and a two-form $X_{\mu\nu}$ we  define the   different projections to flat time and spatial components as follows:
\begin{equation} \label{flatteningrules}
    X_0 \equiv \tau^{\mu} X_{\mu}\,, \qquad X_A \equiv e_{A}{}^{\mu} X_{\mu}\,,\qquad X_{0A} \equiv \tau^{\mu} e_{A}{}^{\nu} X_{\mu\nu}\,, \qquad X_{AB}\equiv e_{A}{}^{ \mu}e_{B}^{\,\, \nu}X_{\mu\nu}\,.
\end{equation}

\subsection{Imposing Constraints} \label{ssec:carrconstr}

As in the relativistic case, the multiplet of gauge fields \eqref{eq:confcarrtrafos} that results from gauging the conformal Carroll algebra, contains more independent fields than are necessary in a gravitational theory. We therefore wish to impose suitable curvature constraints that allow one to solve for some fields in terms of the remaining ones and thus reduce the number of independent fields.

By inspecting the list of curvatures given in \eqref{eq:carrconfcurvs}, one sees that a minimal set of independent fields can be obtained in analogy to the relativistic case, by constraining the curvatures $R_{\mu\nu}(H)$ and $R_{\mu\nu}(P^A)$ of time and spatial translations, as well as the components $R_{\mu B}(J^{AB})$ and $R_{\mu A}(J^{0A})$ of the curvatures of spatial rotations and Carroll boosts. Similar to what we discussed in section \ref{2.2}, one could impose the following constraints:
\begin{align} \label{eq:genCarrconstr}
  R_{\mu\nu}(H) = T_{\mu\nu} \,, \quad R_{\mu\nu}(P^A) = T_{\mu\nu}{}^{A} \,, \qquad \qquad R_{\mu B}(J^{AB}) = 0 \,, \quad R_{\mu A}(J^{0A}) = 0 \,,
\end{align}
where $T_{\mu\nu}$ and $T_{\mu\nu}{}^A$ taken together yield a Carrollian analogue of the torsion tensor. In this paper, we will not consider the case of arbitrary torsion. We will instead closely follow the reasoning that led to general relativity and put as many Carrollian torsion components equal to zero as possible. 
Note however that we should not require that all torsion components vanish. The reason is that some components of $R_{\mu\nu}(P^A)$ do not contain (spin-connection or dilatation) gauge field components that appear algebraically and can be solved for. In our case, this happens for the following $\tfrac12 D(D-1)-1$ components of the spatial translation curvature:
\begin{equation}\label{comma}
R_0{}^{\{A}(P^{B\}})\,,
\end{equation}
where $\{AB\}$ denotes the symmetric traceless part of $AB$. The corresponding torsion components \footnote{The comma in $T_0{}^{\{A,B\}}$ indicates that the two indices to the left of the comma are flat projections of the two curved indices of $T_{\mu\nu}{}^A$.}
\begin{align} \label{eq:intrtorsion}
  T_0{}^{\{A,B\}} \equiv \tau^\mu e^{\{A|\nu} T_{\mu\nu}{}^{|B\}} \,,
\end{align}
define the intrinsic torsion.\,\footnote{Note that we use the term `intrinsic torsion' in a more general sense than is typically done in the mathematical literature, where this term is reserved for the torsion tensor components that do not contain spin-connection components. This is appropriate for space-times with local Carroll symmetries. Since here however, we are dealing with the conformal Carroll algebra, we will for simplicity define the word intrinsic torsion tensor as a torsion tensor that does not contain spin-connection as well as dilatation gauge fields.} They should not be constrained to be zero from the start, since doing so is tantamount to imposing the following geometric constraints
\begin{align}
  R_0{}^{\{A}(P^{B\}}) = 2 \tau^\mu e^{\{A|\nu} \partial_{[\mu} e_{\nu]}{}^{|B\}} = 0 \,,
\end{align}
and is therefore too restrictive. 

The components given in \eqref{eq:intrtorsion} are the only intrinsic torsion ones: all other components of $T_{\mu\nu}{}^A$ and $T_{\mu\nu}$ are not intrinsic. Setting all non-intrinsic torsion components in \eqref{eq:genCarrconstr} equal to zero, we are led to impose the following conventional constraints
\begin{subequations} \label{eq:specconvC}
\begin{alignat}{2}
 R_{\mu\nu}(H) &= 0 \,, \qquad \qquad \qquad  & R_{\mu\nu}(P^A) &= 2\tau_{[\mu} e_{\nu] B}\, T_{0}{}^{\{B,A\}}\,, \label{specialC} \\ R_{\mu B}(J^{AB}) &= 0 \,, \qquad & R_{\mu A}(J^{0A}) &= 0 \,. \label{conventionalC}
\end{alignat}
\end{subequations}
The first two of these (given in \eqref{specialC}) represent in total
\begin{align}
  \frac12 D(D-1) + \frac12 D (D-1)^2 - \frac12 D(D-1) + 1 = \frac12 D^3 - D^2 + \frac12 D + 1 \,,
\end{align}
constraints and can be used to solve for the following equal number of gauge field components: 
\begin{subequations} \label{eq:OmFsolsur}
\begin{align} 
  \omega_\mu{}^{AB} (e,b) &=  2 \, e^{[A|\nu} \, \partial_{[\mu} e_{\nu]}{}^{|B]} - e_{\mu C}\, e^{A\nu}\, e^{B\rho}\, \partial_{[\nu} e_{\rho]}{}^{C} + 2\, e_\mu{}^{[A} e^{B]\nu} b_\nu \,,\label{omAB}\\[.1truecm]
  \omega_0{}^{0A}\left(e,\tau,b\right)&=-2 \tau^{\mu}e^{A \nu}\partial_{[\mu}\tau_{\nu]}+ e^{A\mu} b_\mu \,, \label{om00A}\\[.1truecm]
   \omega^{[A|,0|B]}\left(e,\tau\right)&=-e^{A\mu}e^{B\nu}\partial_{[\mu}\tau_{\nu]}\,, \label{omA0B}\\[.1truecm]
     b_{0}(e,\tau) &=-\frac{2}{(D-1)} \tau^\mu e_A{}^\nu \partial_{[\mu} e_{\nu]}{}^A \equiv-\frac{1}{(D-1)}\mathcal{T}_{0 A,}{}^{A}\,,\label{b}
\end{align}
\end{subequations}
where we have defined $\mathcal{T}_{0 A,}{}^{A}$, corresponding to $T_{0 A,}{}^{A}$ without the $b_{0}$ gauge field. Similarly, one can show that the first conventional constraint of \eqref{conventionalC} can be used to solve for the vector special conformal gauge field $g_\mu{}^A$ as follows:
\begin{equation} \label{gsol}
    g_{\mu}{}^{A}\left(e,\tau,b\right)=-\frac{1}{2(D-3)}\bigg[R^\prime_{\mu B}\left(J^{AB}\right)-\frac{1}{2(D-2)}e_{\mu}{}^{A}R^\prime_{BC}{}\left(J^{BC}\right)-
    \frac{1}{(D-2)}\tau_{\mu}R^\prime_{0B}\left(J^{AB}\right)\bigg]\,.
\end{equation}
Here, $R^\prime_{\mu\nu}(J^{AB})$ is given by the expression \eqref{eq:RJAB} for the curvature tensor $R_{\mu\nu}(J^{AB})$, without the term involving the vector special conformal gauge field $g_\mu{}^A$ and with $\omega_\mu{}^{AB}$ replaced by the expression \eqref{omAB}. Finally, the scalar special conformal gauge field $f_\mu$ can be solved from the second conventional constraint of \eqref{conventionalC} as follows:
\begin{equation} \label{fsol}
    f_{\mu}\left(e,\tau,b\right)=-\frac{1}{2(D-2)}\left[R^\prime_{\mu A}\left(J^{0A}\right)-\frac{\tau_{\mu}}{(D-1)}\left(R^\prime_{0 A}\left(J^{0A}\right)+\frac{1}{2}R^\prime_{AB}\left(J^{AB}\right)\right)\right]\,.
\end{equation}
The primed curvature $R^\prime_{\mu\nu}(J^{0A})$ is given by the expression \eqref{eq:RJ0A} for $R_{\mu\nu}(J^{0A})$ without the terms involving $f_\mu$ and $g_\mu{}^{A}$ and with $\omega_\mu{}^{AB}$ and the dependent parts $\omega_0{}^{0A}$ and $\omega^{[A|,0|B]}$ of the boost connection $\omega_\mu{}^{0A}$ replaced by their expressions \eqref{omAB}, \eqref{om00A} and \eqref{omA0B}. 

Summarizing: the gauge field components
\begin{align}
  \omega_\mu{}^{AB}\,, \qquad \omega_0{}^{0A} \,, \qquad \omega^{[A|,0|B]} \,, \qquad b_0 \,, \qquad g_\mu{}^{A} \,, \qquad f_\mu \,,
\end{align}
are turned into dependent fields by virtue of the conventional constraints \eqref{eq:specconvC}. In what follows, it will be understood that these field components are given by their dependent expressions \eqref{omAB}, \eqref{om00A}, \eqref{omA0B}, \eqref{b}, \eqref{gsol}, \eqref{fsol}, whenever they appear explicitly or implicitly, and we will for simplicity drop the reference to the fields on which they depend. The remaining independent fields are given by:
\begin{align}
  \tau_\mu \,, \qquad e_\mu{}^A \,, \qquad b_A \,, \qquad \omega^{(A|,0|B)} \,.
\end{align}
Two differences with the relativistic case are worth pointing out. First, in the Carroll case it is not possible to solve for all spin-connection components and in particular the boost spin-connection components $\omega^{(A|,0|B)}$ remain independent. The presence of these independent boost connection components is related to the fact that there is intrinsic torsion, see e.g., \cite{Bergshoeff:2023rkk}. Secondly, not all components of the dilatation gauge field are independent; only the spatial components $b_A$ are. Note that the dependent spin-connection fields depend only on $b_A$ and not on the dependent $b_0$ component. 

Let us now comment on how the conformal transformation rules of the independent and dependent fields are changed with respect to \eqref{eq:confcarrtrafos} after imposing the conventional constraints \eqref{eq:specconvC}. First, we consider the homogeneous conformal Carroll transformations. As in the relativistic case, we assume that their action on the independent fields is not modified and thus corresponds to the rules given in \eqref{eq:confcarrtrafos}. To determine how they act on the dependent fields, we first investigate how the conventional constraints \eqref{eq:specconvC} transform under \eqref{eq:confcarrtrafos}. Of the constraints \eqref{specialC}, only the parts
\begin{eqnarray} \label{eq:noninvspecialC}
    R_{BC}(P^A) = 0 \,, \qquad \qquad \qquad R_{0A}(H) = 0 \,,
\end{eqnarray}
that are used to solve for $\omega_{C}{}^{AB}$ and  $\omega_0{}^{0,A}$, are not invariant under the homogeneous conformal Carroll transformations of \eqref{eq:confcarrtrafos}, due to the fact that
\begin{eqnarray}
    \delta_{\mathrm{hom},\, 0} R_{BC}(P_A) &= \lambda^{0}{}_B T_{0\{A,C\}} - \lambda^0{}_{C} T_{0\{A,B\}} \,, \qquad \quad
    \delta_{\mathrm{hom},\, 0} R_{0A}(H) &= -\lambda^0{}_B T_{0}{}^{\{A,B\}} \,.
\end{eqnarray}
As a consequence, the correct homogeneous conformal Carroll transformations of $\omega_{C}{}^{AB}$ and  $\omega_0{}^{0,A}$, that leave \eqref{eq:noninvspecialC} invariant, are of the form
\begin{align}
  \delta_{\mathrm{hom}} \omega_C{}^{AB} = \delta_{\mathrm{hom},\, 0} \omega_C{}^{AB} + \Delta \omega_C{}^{AB} \,, \qquad \qquad \qquad \delta_{\mathrm{hom}} \omega_0{}^{0A} = \delta_{\mathrm{hom},\, 0} \omega_0{}^{0A} + \Delta \omega_0{}^{0A} \,,
\end{align}
where the extra contributions $\Delta \omega_C{}^{AB}$ and $\Delta \omega_0{}^{0A}$ are given by: 
\begin{eqnarray}
\Delta \omega_{C,AB} = \lambda^0{}_{B} T_{0\{A,C\}} - \lambda^{0}{}_A T_{0\{B,C\}} \,, \qquad \qquad \qquad
\Delta \omega_0{}^{0A} = \lambda^0{}_B T_0{}^{\{A,B\}}\,.
\end{eqnarray}
These additional transformations can alternatively be derived by varying the explicit solutions of these dependent spin-connection components. The homogeneous conformal transformation rules of the dependent gauge fields $g_\mu{}^A$ and $f_\mu$ likewise acquire extra terms due to the fact that the conventional constraints \eqref{conventionalC} are not invariant under \eqref{eq:confcarrtrafos} or that their explicit expressions \eqref{gsol}, \eqref{fsol} depend on $\omega_C{}^{AB}$ and $\omega_0{}^{0A}$. We will refrain from giving the modified transformations of all components of $g_\mu{}^A$ and $f_\mu$ and instead only focus on those of $g_{A}{}^{A}$ and $f_0$, as these are the only components that we will need. One finds that the correct transformations of $g_A{}^A$ and $f_0$ take the form
\begin{align}
  \delta_{\mathrm{hom}} g_A{}^A = \delta_{\mathrm{hom},\, 0} g_A{}^A + \Delta g_A{}^A \,, \qquad \qquad \qquad \delta_{\mathrm{hom}} f_0 = \delta_{\mathrm{hom},\, 0} f_0 + \Delta f_0 \,,
\end{align}
where the extra terms $\Delta g_A{}^A$ and $\Delta f_0$ are given by
\begin{eqnarray}
    \Delta g_A{}^A &=& -\frac{1}{2(D-2)} D_A \Delta \omega_B{}^{AB} \,, \\[.1truecm]
    \Delta f_0 &=& \frac{1}{2(D-1)} D_A \Delta \omega_{0}{}^{0A} + \frac{1}{2(D-1)(D-2)} D_A \Delta \omega_B{}^{AB} \,,
\end{eqnarray}
with
\begin{eqnarray}
    D_A \Delta \omega_B{}^{AB} &\equiv& \partial_A \Delta \omega_B{}^{AB} + \omega_A{}^{AD}(e,b) \Delta \omega_{B,D}{}^B -  b_A \Delta \omega_B{}^{AB} \,, \\[.1truecm]
    D_A \Delta \omega_{0}{}^{0A} &\equiv& \partial_A \Delta \omega_{0}{}^{0A} + \omega_A{}^{AB}(e,b) \Delta \omega_{0}{}^{0}{}_{B} - b_A \Delta \omega_{0}{}^{0A} \,.
\end{eqnarray}
It turns out that only the combination $g_A{}^A + f_0$ plays a role in the calculations that follow. Since $\Delta \omega_0{}^{0A} = \Delta \omega_{B}{}^{AB}$, this combination does not receive any modifications in its transformation rule:
\begin{eqnarray}
    \Delta g_A{}^A + \Delta f_0 = 0 \,.
\end{eqnarray}

Next, we turn to how the translations of \eqref{eq:confcarrtrafos} are modified in the presence of the conventional constraints \eqref{eq:specconvC}. Following the same logic as in the relativistic case, we define a time/spatial translation with parameter $\zeta/\zeta^A$ as an infinitesimal diffeomorphism with parameter
\begin{align}
  \xi^\mu = \zeta \tau^\mu + \zeta^A e_A{}^\mu \,,
\end{align}
minus homogeneous conformal Carroll transformations with parameters
\begin{alignat}{2}
  \lambda^{AB} &= \left(\zeta \tau^\mu + \zeta^C e_C{}^\mu\right) \omega_\mu{}^{AB} \,, \qquad &\lambda^{0A} &= \left(\zeta \tau^\mu + \zeta^B e_B{}^\mu\right) \omega_\mu{}^{0A} \,, \nonumber \\
  \lambda_K &= \left(\zeta \tau^\mu + \zeta^A e_A{}^\mu \right) f_\mu \,, \qquad & \lambda_K{}^{A} &= \left(\zeta \tau^\mu + \zeta^B e_B{}^\mu\right) g_\mu{}^A \,, \nonumber \\
  \lambda_D &= \left(\zeta \tau^\mu + \zeta^A e_A{}^\mu\right) b_\mu \,.
\end{alignat}
For the independent fields, this yields the following rules:
\begin{align} \label{eq:transcarrindep}
  \delta_\zeta \tau_\mu &= \tilde{D}_\mu \zeta \,, \qquad \qquad \qquad \delta_\zeta e_\mu{}^A = \tilde{D}_\mu \zeta^A - \zeta_B \tau_\mu T_0{}^{\{B, A\}} + \zeta e_{\mu B} T_0{}^{\{B,A\}} \,,  \nonumber \\
  \delta_\zeta b_A &= - \left(\tilde{D}_A \zeta\right) b_0 - \left(\tilde{D}_A \zeta^B\right) b_B - \zeta b^B T_{0\{A,B\}} - 2 \zeta_B g_A{}^B + \zeta R_{0A}(D)  \nonumber \\ & \qquad + \zeta^B R_{BA}(D) \,, \nonumber \\
  \delta_\zeta \omega^{(A|,0|B)} &= - \left(\tilde{D}^{(A|} \zeta \right) \omega_0{}^{0|B)} - \left(\tilde{D}^{(A|} \zeta^C \right) \omega_C{}^{0|B)} - \frac12 \zeta T_0{}^{\{B,C\}} \omega_C{}^{0A} - \frac12 \zeta T_0{}^{\{A,C\}} \omega_C{}^{0B}  \nonumber \\ & \qquad + 2 \zeta^{(A} f^{B)} - 2 \zeta g^{(A,B)} + \zeta_C R^{C(A|}(J^{0|B)}) + \zeta R_0{}^{(A|}(J^{0|B)}) \,.
\end{align}
We will refrain from giving the translation rules of the dependent fields. They can be found using the same prescription, or by varying their explicit expressions under \eqref{eq:transcarrindep}. 

Finally, we mention the following useful formulae for later use:
\begin{eqnarray}
e_A{}^\mu g_\mu{}^A &=&-\frac{1}{4(D-2)}R^\prime_{AB}(J^{AB})\,,\\[.1truecm]
\tau^\mu f_\mu &=& -\frac{1}{2(D-1)} R^\prime_{0A}(J^{0A}) + \frac{1}{4(D-1)(D-2)}R^\prime_{AB}(J^{AB})\,,\\[.1truecm]
e_A{}^\mu g_\mu{}^A + \tau^\mu f_\mu &=& -\frac{1}{4(D-1)}\left[ R^\prime_{AB}(J^{AB}) + 2 R^\prime_{0A}(J^{0A})\right]\,.
\end{eqnarray}

\subsection{Carroll Gravity from a Compensating Mechanism} \label{ssec:carrgravs}

In this subsection we will couple the conformal Carroll gravity theory that we constructed in the previous subsection to a massless  electric and magnetic Carroll scalar field, respectively. Furthermore, we will show how, upon gauge-fixing, this leads to a special electric Carroll  gravity theory and a magnetic Carroll gravity theory. We first discuss the electric case.
\vskip .2truecm

\noindent {\bf The electric case.}\ \  Our starting point is the following Lagrangian for a massless  electric Carroll scalar field:
\begin{equation}
    \mathcal{L}_{\text{electric scalar}}=-\frac{1}{2}\phi\partial_{t}\partial_{t}\phi\,. \label{elecscal}
\end{equation}
As in the relativistic case, we will render this action invariant under local homogeneous conformal Carroll transformations (as well as, manifestly, under diffeomorphisms. We assume that the transformation rule of the scalar field $\phi$ under  homogeneous conformal Carroll transformations is given by a dilatation with weight $w$ as follows
\begin{align}
  \label{eq:dilphiUR}
  \delta_{\text{hom}} \phi = w\, \lambda_D \, \phi \,.
\end{align}
As a first step we replace the time derivative $\partial_t\phi$ of $\phi$ by a covariant derivative $D_0\phi\,:$
\begin{equation}\label{covt}
    \partial_t\phi  \rightarrow D_{0}\phi \equiv \tau^{\mu} \left(\partial_\mu - w\, b_\mu \right) \phi \,.
\end{equation}
Note that the second term in the above covariant derivative is related to  the transformation rule \eqref{eq:dilphiUR} by replacing the dilatation parameter $\lambda_D$ by the corresponding gauge field $b_\mu$. In a second step we consider the transformation rule of the covariant derivative $D_0\phi$ under the homogeneous conformal Carroll transformations:
\begin{align}\label{transfcovt}
  \delta_{\text{hom}}\left(D_{0} \phi\right) =  (w + 1)\, \lambda_D \, D_{0} \phi \,.
\end{align}
This transformation rule shows that the second covariant time derivative $D_0D_0\phi$ of $\phi$ is given by
\begin{align}
  \label{eq:cbox}
  D_{0}D_{0} \phi \equiv \tau^{\mu} \left[\partial_\mu\left(D_{0} \phi\right) - (w+1)\, b_\mu \, D_{0} \phi \right] \,.
\end{align}
  Using the transformation rules \eqref{inverseV}  of $\tau^\mu$ and the transformation rule \eqref{eq:confcarrtrafos} of $b_\mu$ we derive that
\begin{align} \label{eq:trafocbox2}
\delta_{\text{hom}}( D_{0}D_{0} \phi) = (w + 2) \, \lambda_D \,  D_{0}D_{0} \phi \,.
\end{align}

We now consider the following Lagrangian describing the coupling of the electric Carroll scalar to conformal Carroll gauge fields:
\begin{equation}
    \mathcal{L}_{\rm electric\ coupling}=-\frac{1}{2}e\phi D_{0}D_{0} \phi\,. \label{Lag}
\end{equation}
Here, $e=\text{det}(\tau_{\mu},e_{\mu}{}^{A})$  which transforms under dilatations as
\begin{equation}\label{transfe}
\delta_{\text{hom}} e = -D\lambda_D e\,.
\end{equation}
Combining the transformation rules \eqref{eq:dilphiUR}, \eqref{eq:trafocbox2} with \eqref{transfe} we find that the Lagrangian \eqref{Lag} transforms as
\begin{equation}
   \delta_{\text{hom}} \mathcal{L}_{\rm electric\ coupling}=-\frac{1}{2}[-D+2w+2]\lambda_D\, e\phi D_{0}D_{0} \phi\,.
\end{equation}
This shows that the proposed Lagrangian is invariant under dilatations provided we take the scaling weight $w$ to be
\begin{equation}
     w=\frac{D-2}{2}\,.
\end{equation}

Having constructed an invariant Lagrangian we can now gauge-fix the dilatations by imposing the gauge condition $\phi=1$ and obtain the following Lagrangian:
\begin{equation}
    \mathcal{L}_{\rm electric\ Carroll}= \frac{w^2}{2} e\, b_0^{2}\,.
\end{equation}
In deriving this, we partially integrated and used the identity
\begin{equation} \label{detau}
    \partial_{\mu}\left(e \tau^{\mu} \right) = 2\, e\, \tau^{\mu} e_{A}{}^{\nu} \partial_{[\mu} e_{\nu]}{}^{A} = - (D - 1)\, e\, b_0 \,,
\end{equation}
where the last equality follows from the conventional constraint $R_{0A}(P^A) = 0$.
Substituting the expression for the dependent dilatation gauge field given in \eqref{b} we find a Lagrangian describing a particular form of electric Carroll gravity:\,\footnote{We use here a calligraphic notation to indicate that $\mathcal{T}_{0 A,}{}^{A}$ is an intrinsic torsion tensor with respect to the Carroll algebra but not with respect to the conformal Carroll algebra.}
\begin{equation}
\mathcal{L}_{\rm electric\ Carroll}=\frac{w^2}{2(D-1)^{2}}\,e \,\mathcal{T}_{0 A,}{}^{A}\mathcal{T}_{0 B,}{}^{B}\,.
\end{equation}
Note that this is not the most general electric Carroll gravity theory given in the literature \cite{Henneaux:1979vn}. There exists  also a conformal electric Carroll gravity theory given by
\begin{equation}
\mathcal{L}_{\rm conformal\ Carroll} \sim \,e \, T_0{}^{\{A,B\}}T_{0\{A,B\}} \,,
\end{equation}
that transforms homogeneously  under local dilatations (but does not contain a dilatation gauge field). This conformal Carroll gravity Lagrangian is an example of a Carroll gravity that cannot be obtained starting from a dynamical matter Lagrangian.
\vskip .5truecm

\noindent {\bf The magnetic case.} In this case, the Lagrangian that we wish to make invariant under local homogeneous conformal Carroll transformations (and diffeomorphisms) is that of a massless magnetic Carroll scalar \cite{Henneaux:2021yzg,deBoer:2021jej}:
\begin{equation}
\mathcal{L}_{\mathrm{magnetic\ scalar}}=\pi\partial_{t}\phi-\frac{1}{2}\partial^{A}\phi\partial_{A}\phi\,, \label{magscal2}
\end{equation}
where $\pi$ is an independent Lagrange multiplier field.
This Lagrangian is invariant under the constant Carroll boost transformations
\begin{equation}
\delta\pi = \lambda^{0A}\partial_A\phi\,,\hskip 2truecm \delta \partial_A\phi = \lambda^0{}_A\partial_t\phi \,,
\end{equation}
and under the constant transformation
\begin{equation}\label{constantK}
\delta \pi = \alpha\lambda_K \phi \,,
\end{equation}
for any $\alpha$. 

As before, we assume that the scalar field $\phi$ transforms under local homogeneous conformal Carroll transformations with a dilatation with scaling weight $w$. We already explained in the electric case how the time derivative $\partial_t\phi$ of $\phi$ can be extended to the covariant derivative $D_0\phi$ defined in eq.~\eqref{covt} and how this covariant derivative transforms according to eq.~\eqref{transfcovt} under the homogeneous Carroll transformations. Similarly, we extend the spatial derivative $\partial_A\phi$ of $\phi$ to the following covariant derivative:
\begin{align}
  \label{eq:DphiUR2}
  D_{A} \phi&\equiv  e_{A}{}^{\mu} \left(\partial_\mu - w\, b_\mu \right) \phi \,.
\end{align}
We find that under the homogeneous conformal Carroll transformations this covariant derivative transforms as follows:
\begin{align}
  \delta_{\mathrm{hom}} \left(D_{A} \phi\right) = - \lambda_{A}{}^{B} \, D_{B} \phi + \lambda^{0}{}_{A} \, D_{0} \phi + (w + 1)\, \lambda_D \, D_{A} \phi - 2\, w \, \lambda_{K A} \, \phi \,.
\end{align}

We now propose the following Ansatz Lagrangian describing the coupling of a massless magnetic Carroll scalar to conformal Carroll gauge fields:
\begin{equation} 
\mathcal{L}_{\text{Ansatz}}=e\pi D_{0}\phi-\frac{1}{2}e D^{A}\phi D_{A}\phi\,. \label{magscal}
\end{equation}
This Ansatz Lagrangian is invariant under spatial rotations. Requiring it to be invariant under dilatations and Carroll boosts requires us to take
\begin{equation}
\delta \pi = \lambda^{0A}D_A\phi + \tfrac12 D\lambda_D\pi\,\hskip 1truecm \textrm{and}\hskip 1truecm w = \tfrac12 (D-2)\,.
\end{equation}

We next consider the vector special conformal transformations. We find that the Ansatz lagrangian is not invariant under these transformations:
\begin{equation}
     \delta \mathcal{L}_{\rm Ansatz}=2 w e\lambda_{K}{}^{A} \, \phi D_A\phi\,.
\end{equation}
To cancel this term, we perform a partial differentiation. Using the identity
\begin{equation}
\partial_\mu \big(e\, e_{A}{}^\mu\big) = 2\, e\, e_{A}{}^{\mu} e_{B}{}^\nu \partial_{[\mu} e_{\nu]}{}^B  - 2\,e\,\tau^\mu e_{A}{}^{\nu} \partial_{[\mu}\tau_{\nu]} \,,
\end{equation}
together with the conventional constraints
\begin{equation}
R_{AB}(P^B) = R_{0A}(H) =0 \,,
\end{equation}
we find that
\begin{align}
   & \delta \mathcal{L}_{\rm Ansatz} = -we\, \omega_{0,}{}^{0A}\lambda_{K A}\phi^2 - we\,e_A{}^\mu\left(D_\mu\lambda_K{}^A\right)\phi^2\,, \\[10pt]
   & \text{with} \qquad \qquad e_A{}^{\mu} \left(D_{\mu} \lambda_K{}^A\right) \equiv e_{A}{}^{\mu} \left(\partial_{\mu} \lambda_K{}^A + \omega_{\mu}{}^{A}{}_{B} \lambda_K{}^{B} - \lambda_K{}^{A} b_{\mu} \right) \nonumber \,.
\end{align}
 Both terms can be canceled by adding two terms that involve the gauge fields of the scalar and vector special conformal transformation which, remarkably, does not break the boost symmetry. Doing this, we end up with the following modified Lagrangian describing the coupling of a massless magnetic Carroll scalar to conformal Carroll gauge fields:
\begin{align}
  \label{eq:Lmag1}
  \mathcal{L}_{\rm magnetic\ coupling} &= e \pi D_0 \phi -\frac{e}{2} D^A \phi D_A \phi+w e g_A{}^A \phi^2+w e f_0 \phi^2\,.
\end{align}
This Lagrangian is invariant under spatial rotations, boosts, dilatations and the vector special conformal transformations.

We still need to verify that the Lagrangian \eqref{eq:Lmag1} is also invariant under the  scalar special conformal transformations. Allowing a $K$-variation $\delta_K \pi$ of the Lagrange multiplier $\pi$, we find
\begin{align}
   & \delta_K \mathcal{L} _{\rm magnetic\ coupling }=e \delta_K \pi D_0 \phi +we\, \big(D_0\lambda_K \big)\phi^2\,, \\[10pt]
   & \text{with} \qquad D_0 \lambda_K \equiv \tau^{\mu} \left( \partial_{\mu} \lambda_K  - \lambda_K b_{\mu} \right) \,. \nonumber
\end{align}
We now partially differentiate the covariant time derivative in the second term. Using \eqref{detau}, 
we obtain invariance under local $K$-transformations provided we take
\begin{equation}
\delta_K \pi = (D-2)\,\lambda_K\phi\,.
\end{equation}
This local $K$-transformation reduces to the global $K$-transformation \eqref{constantK} for $\alpha=D-2$.

We now fix the dilatations by imposing $\phi=1$. The invariance of the Lagrangian \eqref{eq:Lmag1} under the vector special conformal transformations combined with the fact that all dependent gauge fields transform under these transformations due to their dependence on $b_A$ guarantees that all $b_A$-dependent terms will cancel out. One is then left with the following Lagrangian for magnetic Carroll gravity:
 \begin{equation}\label{mCarrollgr}
\mathcal{L}_{\rm magnetic\ Carroll} = \frac{1}{2} \frac{D-2}{D-1}e\, \left\{\pi\, \mathcal{T}_{0A,}{}^A
-\frac{1}{4}\left[ R^\prime_{AB}(J^{AB})(e,\tau) + 2 R^\prime_{0A}(J^{0A})(e,\tau)\right]\right\}\,,
\end{equation}
where $R^\prime(e,\tau)$ are curvature tensors with respect to the (non-conformal) Carroll algebra. At first sight, this expression differs from the result for magnetic Carroll gravity given in \cite{Bergshoeff:2017btm} due to the presence of the first term. However, as pointed out in \cite{Bergshoeff:2023rkk}, the curvature terms contain many more Lagrange multipliers. These are the independent spin-connection components $\omega^{(A,0B)}$ that impose the geometric constraints: 
\begin{equation}
\mathcal{T}_0{}^{(A,B)} = 0\,.
\end{equation}
In particular, this implies that  the first term in \eqref{mCarrollgr} can be absorbed by a redefinition of the Lagrange multiplier field $\omega_{A,}{}^{0A}$. This is supported by the fact that the Lagrangian \eqref{mCarrollgr} is invariant under the $K$-transformation:
\begin{equation}
\delta\pi = (D-2)\,\lambda_K\,,\hskip 2truecm \delta \omega_A,{}^{0A} = -2\lambda_K \,,
\end{equation}
that can be gauge-fixed by setting $\pi=0$. We thus end up with the following Lagrangian for magnetic Carroll gravity \cite{Bergshoeff:2017btm}:
\begin{equation}\label{mCarrollgr2}
\mathcal{L}_{\rm magnetic\ Carroll} = -\frac{1}{8} \frac{D-2}{D-1}e\, \left[ R^\prime_{AB}(J^{AB})(e,\tau) + 2 R^\prime_{0A}(J^{0A})(e,\tau)\right]\,.
\end{equation}
Alternatively, one could gauge-fix the $K$-transformations by imposing $\omega_A,{}^{0A}=0$. In that case, due to compensating transformations, the Lagrange multiplier field $\pi$  would behave as a special boost spin-connection component.

We note that gauge-fixing a similar Lagrangian that was derived  from a Carroll expansion of general relativity in \cite{Hansen:2021fxi}, gives the same magnetic Carroll gravity theory with an additional term given by 
\begin{equation}
\mathcal{L}_{\rm extra} = \chi_{AB}\, T_0{}^{\{A,B\}}\,,
\end{equation}
where $\chi_{AB}$ is a Lagrange multiplier field symmetric and traceless in $A$ and $B$. This extra term is another example of a Carroll gravity invariant that can not be obtained from a dynamical matter Lagrangian.

\section{The curious case of Galilei Gravity}

Similar to the Carroll case, there exist two Galilei gravity theories at the two-derivative level. The Vielbeine $(\tau_{\mu}, e_{\mu}{}^{A})$ (with $A=1,\cdots, D$) that appear in these theories transform under local spatial rotations (with parameters $\lambda^{AB} = -\lambda^{BA}$) and local Galilean boosts (with parameters $\lambda^{0A}$) as
\begin{eqnarray}
    \delta \tau_{\mu} = 0 \,, \qquad \qquad \qquad \delta e_{\mu}{}^{A} = - \lambda^{A}{}_{B} e_{\mu}{}^{B} - \lambda^{0A} \tau_{\mu} \,.
\end{eqnarray}
One also introduces dual Vielbeine $(\tau^{\mu}, e_A{}^{\mu})$ via the relations \eqref{invVielbdef}. Their transformation rules under spatial rotations and Galilean boosts are given by
\begin{eqnarray}
\delta \tau^{\mu} = \lambda^{0A} e_A{}^{\mu} \,, \qquad \qquad \qquad \delta e_A{}^{\mu} = -\lambda_{A}{}^{B} e_B{}^{\mu} \,.
\end{eqnarray}
The rules \eqref{flatteningrules} can then again be used to turn curved ($\mu$) indices into flat ones ($0$ and $A$). The first Galilei gravity theory appears as the leading order term in a $1/c^2$--expansion of general relativity, similar to electric Carroll gravity. For this reason, we will refer to it as `electric Galilei gravity'. Its action is proportional to
\begin{eqnarray} \label{elGalgrav}
    \int \rmd^D x \, e \, \tau^{AB} \tau_{AB} \,,
\end{eqnarray}
where $\tau_{\mu\nu} = 2 \partial_{[\mu} \tau_{\nu]}$. Likewise, the second Galilei gravity theory corresponds to the subleading term in a $1/c^2$--expansion of general relativity. Since this is analogous to magnetic Carroll gravity, we will refer to it as `magnetic' Galilei gravity. Its action is proportional to
\begin{eqnarray} \label{magnGalgrav}
    \int \rmd^D x \, e \, \left[e_A{}^{\mu} e_B{}^{\nu} R_{\mu\nu}(J^{AB}) + A^{AB} \tau_{AB} \right] \,.
\end{eqnarray}
Here, $A_{AB}$ is a Lagrange multiplier field that imposes the constraint $\tau_{AB} = 0$ and 
\begin{eqnarray} 
    R_{\mu\nu}(J^{AB}) &\equiv& 2 \partial_{[\mu} \omega_{\nu]}{}^{AB}(e,\tau) + 2 \omega_{[\mu|}{}^{A}{}_{C}(e,\tau) \omega_{|\nu]}{}^{CB}(e,\tau) \,, \qquad \text{with} \\[.1truecm]
    \omega_{\mu}{}^{AB}(e,\tau) &\equiv& 2 e^{[A|\nu} \partial_{[\mu} e_{\nu]}{}^{|B]} - e_{\mu C} e^{A\nu} e^{B\rho} \partial_{[\nu} e_{\rho]}{}^{C} - \frac{2}{D-3} e_{\mu}{}^{[A} \tau_{0}{}^{B]} \,.
\end{eqnarray}

Above, we saw that both electric and magnetic Carroll gravity can be obtained by applying the conformal approach to the conformal Carroll algebra and compensating electric and magnetic Carroll scalars. It is then natural to ask whether electric and magnetic Galilei gravity can be obtained in an analogous manner, starting from the Galilean conformal algebra, the Galilean counterpart of the conformal Carroll algebra \eqref{eq:CCA1}. In this regard, it is useful to note that both electric and magnetic Galilei gravity are already invariant under local scale symmetries. These are however anisotropic rescalings (with parameter $\lambda$), given by:
\begin{eqnarray}
    & & \text{for electric Galilei gravity: } \ \ \qquad \delta \tau_{\mu} =  \tfrac{1}{3} (D - 5) \lambda \tau_{\mu} \,, \qquad \qquad \delta e_{\mu}{}^{A} = -\lambda e_{\mu}{}^{A} \,, \\[.1truecm]
    & & \text{for magnetic Galilei gravity: } \ \ \, \quad \delta \tau_{\mu} =  (D-3) \lambda \tau_{\mu} \,, \qquad \quad \, \delta e_{\mu}{}^{A} = -\lambda e_{\mu}{}^{A} \,, \nonumber \\[.1truecm] & & \qquad \qquad \qquad \qquad \qquad \qquad \quad \ \, \delta A_{AB} = -(D-3) \lambda A_{AB} \,.
\end{eqnarray}
Neither electric nor magnetic Galilei gravity are invariant under the isotropic rescalings
\begin{eqnarray} \label{dilGal}
    \delta \tau_{\mu} = - \lambda_D \tau_{\mu} \,, \qquad \qquad \qquad \delta e_{\mu}{}^{A} = -\lambda_D e_{\mu}{}^{A} \,,
\end{eqnarray}
that constitute the dilatation symmetry of the Galilean conformal algebra. The Lagrangian of electric Carroll gravity does however transform covariantly, i.e., without a derivative of the parameter, under such isotropic local dilatations:  
\begin{eqnarray} 
    \delta \left(e \tau^{AB} \tau_{AB} \right) = - (D-2) \lambda_D \, e \tau^{AB} \tau_{AB} \,.
\end{eqnarray}
This implies in particular that the electric Galilei action \eqref{elGalgrav} can not be obtained by applying the conformal approach to the Galilean conformal algebra and a \emph{dynamical} compensating scalar. Indeed, in order to render the electric Galilei action \eqref{elGalgrav} invariant under the Galilean conformal algebra, and in particular the dilatations \eqref{dilGal}, it suffices to couple it to a non-dynamical compensating scalar $\phi$ with dilatation weight $w = (D-2)/2$ as follows 
\begin{eqnarray} \label{elGalgravconf}
    \int \rmd^D x \, e \, \phi^2 \tau^{AB} \tau_{AB} \,.
\end{eqnarray}
The electric Galilei gravity action \eqref{elGalgrav} is then retrieved as usual by imposing the dilatation gauge fixing $\phi = 1$.

The magnetic Galilei action \eqref{magnGalgrav} does not simply rescale covariantly under the Galilean conformal dilatations \eqref{dilGal}. In particular, applying such an isotropic dilatation to the first term of \eqref{magnGalgrav}, leads to $\partial_A \lambda_D \partial^A \lambda_D$ terms. While this suggests that it might be possible to obtain magnetic Galilei gravity from the conformal approach applied to the Galilean conformal algebra and a compensating magnetic Galilei scalar with action
\begin{eqnarray} \label{eq:flatmagnGalscalar}
  \int \rmd^D x \left(-\frac12 \partial_A \phi \partial^A \phi\right) \,,
\end{eqnarray}
we will now show that this is not the case, by explicitly going through the procedure.

Like the conformal Carroll algebra, the Galilean conformal algebra is an In\"on\"u-Wigner contraction of the relativistic conformal algebra \eqref{eq:relconfalg}. This contraction is obtained by splitting the index $\hat{A}$ into 0 and $A = 1,\cdots,D-1$, and subsequently redefining the generators of \eqref{eq:relconfalg} with a contraction parameter $c$ as follows
\begin{alignat}{3}
  M_{0A} \ \ &\rightarrow \ \ c\, G_{A} \,, \qquad \qquad & M_{AB} \ \ &\rightarrow \ \ J_{AB} \,, \qquad \qquad & P_0 \ \ &\rightarrow \ \ c^{-1} H \,, \qquad \qquad P_A \ \ \rightarrow \ \ P_A \,, \nonumber \\
  K_0 \ \ &\rightarrow \ \ c \, K \,, \qquad \qquad &K_{A} \ \ &\rightarrow \ \ c^2 \, K_{A} \,, \qquad \qquad & D \ \ &\rightarrow \ \ D \,,
\end{alignat}
and by finally taking the $c \rightarrow \infty$ limit. The non-zero commutation relations of the resulting Galilean conformal algebra are then given by (see e.g., \cite{Bagchi:2009my})
\begin{alignat}{2} \label{eq:galconfalg}
  \comm{J_{AB}}{J_{CD}} &= 4 \delta_{[A[C} J_{D]B]} \,, \qquad \qquad & \comm{J_{AB}}{G_{C}} &= -2 \delta_{C[A} G_{B]} \,, \nonumber \\
  \comm{H}{G_{A}} &= -P_A \,, \qquad \qquad & \comm{K}{G_A} &= -K_A \,, \nonumber \\
  \comm{J_{AB}}{P_C} &= -2  \delta_{C[A} P_{B]} \,, \qquad \qquad & \comm{J_{AB}}{K_{C}} &= -2 \delta_{C[A} K_{B]} \,, \nonumber \\
  \comm{H}{K} &= -2 D\,, \qquad \qquad & \comm{D}{H} &= H \,, \nonumber \\
  \comm{H}{K_A} &= 2 G_{A} \,, \qquad \qquad & \comm{D}{P_{A}} &= P_A \,, \nonumber \\
  \comm{P_A}{K} &= -2 G_{A} \,, \qquad \qquad & \comm{D}{K} &= -K \,, \nonumber \\
  \comm{D}{K_A} &= - K_A \,.
\end{alignat}
In what follows we will be mostly interested in the homogeneous Galilean conformal transformations, that consist of spatial rotations $J_{AB}$, Galilean boosts $G_A$, dilatations $D$ and the special conformal transformations $K$ and $K_A$.

Introducing gauge fields $\tau_\mu$, $e_\mu{}^A$, $\omega_\mu{}^{AB}$, $\omega_\mu{}^{0A}$, $b_\mu$, $f_\mu$ and $g_\mu{}^A$ (associated to $H$, $P_A$, $J_{AB}$, $G_A$, $D$, $K$ and $K_A$ respectively), the commutation relations \eqref{eq:galconfalg} give the following homogeneous Galilean conformal transformation rules:\,\footnote{Strictly speaking, we should denote these transformations as $\delta_{\mathrm{hom},\, 0}$ to be in line with the notation of the previous sections. Here however, we will not go through all details of the conformal approach, but only focus on those aspects that are relevant for the case at hand. For our purposes, it will not be necessary to distinguish between e.g., $\delta_{\mathrm{hom},\, 0}$ and $\delta_{\mathrm{hom}}$ and we will simply use $\delta$ everywhere.}
\begin{align}
  \label{eq:galconftrafos}
  \delta \tau_\mu &=  -\lambda_D \tau_\mu \,, \nonumber \\
  \delta e_\mu{}^A &=  -\lambda^{A}{}_B  e_{\mu}{}^{B} - \lambda^{0A}  \tau_\mu - \lambda_D e_\mu{}^A \,, \nonumber \\
  \delta \omega_\mu{}^{AB} &= \partial_\mu \lambda^{AB} - 2 \lambda^{[A|C|}  \omega_{\mu C}{}^{B]} \,, \nonumber \\
  \delta \omega_\mu{}^{0A} &= \partial_\mu \lambda^{0A} - \lambda^{A}{}_{B} \omega_\mu{}^{0B} - \lambda^{0B} \omega_{\mu B}{}^A + 2 \lambda_K{}^A  \tau_\mu - 2 \lambda_K  e_\mu{}^A \,, \nonumber \\
  \delta b_\mu &= \partial_\mu \lambda_D - 2 \lambda_K  \tau_\mu \,, \nonumber \\
  \delta f_\mu &= \partial_\mu \lambda_K + \lambda_D f_\mu - \lambda_K  b_\mu \,, \nonumber \\
  \delta g_\mu{}^A &= \partial_\mu \lambda_K{}^A - \lambda_K{}^B  \omega_{\mu B}{}^A - \lambda^A{}_B g_\mu{}^B + \lambda_K \omega_\mu{}^A - \lambda^{0A}  f_\mu - \lambda_K{}^{A} b_\mu + \lambda_D  g_{\mu}{}^A \,.
\end{align}
Here, $\lambda^{AB}$, $\lambda^{0A}$, $\lambda_D$, $\lambda_K$ and $\lambda_K{}^A$ are the parameters corresponding to the generators $J_{AB}$, $G_A$, $D$, $K$ and $K_A$ respectively. Note that the dual Vielbeine $\tau^{\mu}$ and $e_{A}{}^{\mu}$ then transform as follows:
\begin{align} \label{eq:invtetrafo}
  \delta \tau^\mu &= \lambda^{0A}  e_{A}{}^{\mu} + \lambda_D \tau^\mu \,, \qquad \qquad \delta e_{A}{}^{\mu} = -\lambda_A{}^B  e_{B}{}^{\mu} + \lambda_D  e_{A}{}^{\mu} \,.
\end{align}
The curvatures that are covariant with respect to the transformations \eqref{eq:galconftrafos} read
\begin{align}
  \label{eq:galconfcurvs}
  R_{\mu\nu}(H) &= 2 \partial_{[\mu} \tau_{\nu]} + 2  b_{[\mu} \tau_{\nu]} \,, \nonumber \\
  R_{\mu\nu}(P^A) &= 2 \partial_{[\mu} e_{\nu]}{}^A + 2 \omega_{[\mu}{}^{AB}  e_{\nu]B} + 2  \omega_{[\mu}{}^{0A}  \tau_{\nu]} + 2  b_{[\mu} e_{\nu]}{}^A \,, \nonumber \\
  R_{\mu\nu}(J^{AB}) &= 2 \partial_{[\mu} \omega_{\nu]}{}^{AB} + 2  \omega_{[\mu}{}^{[A|C|} \omega_{\nu]C}{}^{B]} \,, \nonumber \\
  R_{\mu\nu}(G^A) &= 2 \partial_{[\mu} \omega_{\nu]}{}^{0A} + 2  \omega_{[\mu}{}^{AB}  \omega_{\nu]}{}^{0}{}_{B} - 4 g_{[\mu}{}^A  \tau_{\nu]} + 4  f_{[\mu} e_{\nu]}{}^A \,, \nonumber \\
  R_{\mu\nu}(D) &= 2 \partial_{[\mu} b_{\nu]} + 4  f_{[\mu}  \tau_{\nu]} \,, \nonumber \\
  R_{\mu\nu}(K) &= 2 \partial_{[\mu} f_{\nu]} + 2 f_{[\mu}  b_{\nu]} \,, \nonumber \\
  R_{\mu\nu}(K^A) &= 2 \partial_{[\mu} g_{\nu]}{}^A + 2 g_{[\mu}{}^B  \omega_{\nu]B}{}^A - 2 f_{[\mu} \omega_{\nu]}{}^{0A} + 2 g_{[\mu}{}^A  b_{\nu]} \,.
\end{align}
Note that in contrast to the Carroll case, the conformal Galilean curvature $R_{\mu\nu}(J^{AB})$ does not contain any component of a gauge field other than the spatial spin-connection $\omega_{\mu}{}^{AB}$. This already indicates that the term $e e_{A}{}^{\mu} e_{B}{}^{\nu} R_{\mu\nu}(J^{AB})$ of magnetic Galilei gravity can not appear as a component of a gauge field that becomes dependent in applying the conformal approach to the Galilean conformal algebra.

In order to see what theory we do get from this approach, when applied to the flat space magnetic Galilei scalar Lagrangian \eqref{eq:flatmagnGalscalar}, we promote the latter to a curved space Lagrangian that is invariant under local homogeneous Galilean conformal transformations. In doing this, we assume that the compensating scalar $\phi$ only transforms with weight $w$ under local dilatations:
\begin{align}
  \label{eq:dilphi}
  \delta \phi = w \lambda_D \phi \,.
\end{align}
The proper covariantization of $\partial_A \phi$ is then
\begin{align} \label{eq:Daphi}
  D_A \phi \equiv \left(\partial_A - w b_A(e,\tau) \right) \phi \,.
\end{align}
Here, we have written the (spatial part of the) dilatation gauge field as $b_A(e,\tau)$ to indicate that this is a dependent gauge field, as it can be solved from the conventional constraint
\begin{align} \label{eq:RHconstrGal}
  R_{0A}(H) \equiv 0 \,.
\end{align}
Explicitly, one finds the following expression for $b_A$ from this constraint:
\begin{align} \label{solbAGal}
  b_A(e,\tau) = \tau_{0A} \,.
\end{align}
Under homogeneous Galilean conformal transformations $b_A(e,\tau)$ transforms as
\begin{align} \label{eq:dpba}
  \delta b_A(e,\tau) &= \partial_A \lambda_D + \lambda_D b_A(e,\tau) - \lambda_A{}^B \, b_B(e,\tau) + \lambda^{0B}\, \tau_{BA} \,.
\end{align}
As can be seen from \eqref{eq:galconftrafos}, the last term does not follow from the Galilean conformal algebra. The presence of this extra boost transformation is a consequence of the fact that the constraint \eqref{eq:RHconstrGal} is not invariant under the transformation rules \eqref{eq:galconftrafos}, but rather one has
\begin{align}
  \delta R_{0A}(H) = \lambda^{0B} R_{BA}(H) = \lambda^{0B} \tau_{BA} \,.
\end{align}

Under homogeneous Galilean conformal transformations, $D_A\phi$ transforms as
\begin{align} \label{eq:trafoDaphi}
  \delta D_A \phi = -\lambda_A{}^B  D_B \phi + (w+1) \lambda_D D_A \phi - w  \lambda^{0B} \tau_{BA} \phi \,.
\end{align}
The appropriate generalization of \eqref{eq:flatmagnGalscalar} that we wish to consider is then given by:
\begin{align} \label{eq:Lmagn0}
  \mathcal{L}_{\mathrm{magn,\,} 0} &=  - \frac{e}{2} D^A \phi D_A \phi \,,
\end{align}
where $e = \mathrm{det}(\tau_\mu,e_\mu{}^A)$. Since
\begin{align}
  \delta e = -D \lambda_D e \,,
\end{align}
one finds that the Lagrangian \eqref{eq:Lmagn0} is dilatation invariant if the weight $w$ is chosen to satisfy
\begin{align} \label{eq:weightcond}
 w = \frac{D}{2} - 1 \,.
\end{align}
Assuming this holds, one finds that under Galilean boosts
\begin{align}
  \delta \mathcal{L}_{\mathrm{magn,\,} 0} &= w e \lambda^{0A}  \left(D^B\phi\right)  \tau_{AB} \phi = \frac{e}{2}  w  \lambda^{0A} D^B (\phi^2) \tau_{AB}\,.
\end{align}
Note that this situation is different from the magnetic Carroll case, where the non-invariance of the (boost-invariant) Ansatz Lagrangian \eqref{magscal} under vector special conformal transformations could be cured by adding a specific combination of special conformal gauge field components that is still invariant under Carroll boosts. Here, we no longer have invariance under boosts and in order to solve this, we introduce and extra Lagrange mulitplier field $\pi_{AB} = \pi_{[AB]}$ that transforms as
\begin{align} \label{eq:pitrafo}
  \delta \pi_{AB} &= -\frac{w}{2} \lambda^0{}_{[A} D_{B]}(\phi^2) - 2 \lambda_{[A}{}^C  \pi_{|C|B]} + (2w +1) \lambda_D \pi_{AB} \,,
\end{align}
and consider the Lagrangian
\begin{align} \label{eq:Lmagn}
  \mathcal{L}_{\mathrm{magn}} &= \mathcal{L}_{\mathrm{magn,\,} 0} + e \pi^{AB} \tau_{AB} = -\frac{e}{2} D^A \phi D_A \phi + e \pi^{AB} \tau_{AB} \,.
\end{align}
Since $e$ and $\tau_{AB}$ have dilatation weight $-D$ and 1 respectively, the last term is dilatation invariant by virtue of \eqref{eq:weightcond}. The boost transformation of $\pi_{AB}$ has moreover been chosen such that $\mathcal{L}_{\mathrm{magn}}$ is boost invariant.

To pass to a Lagrangian that is invariant under only homogeneous Galilean transformations, we fix the superfluous conformal ones. The only one that acts non-trivially on the fields in $\mathcal{L}_{\mathrm{magn}}$ is the dilatation symmetry and it is easily fixed by imposing the gauge
\begin{align}
  \phi = 1 \,.
\end{align}
With this gauge fixing condition, the Lagrangian $\mathcal{L}_{\mathrm{magn}}$ becomes
\begin{align}
  \label{eq:Lmagnfix}
  \mathcal{L}_{\mathrm{magn,\, fixed}} &=  -\frac{e}{2}  w^2  b^A(e,\tau) b_{A}(e,\tau) + e \pi^{AB}  \tau_{AB} = -\frac{e}{2} w^2 \tau_0{}^A \tau_{0A} + e \pi^{AB}  \tau_{AB} \,,
\end{align}
while the transformation rule \eqref{eq:pitrafo} becomes
\begin{align}
  \delta^\prime \pi_{AB} &= w^2 \lambda_{[A} b_{B]}(e,\tau) - 2 \lambda_{[A}{}^C  \pi_{|C|B]} = w^2  \lambda_{[A}  \tau_{|0|B]} - 2 \lambda_{[A}{}^C  \pi_{|C|B]} \,. 
\end{align}
One thus sees that, even though the conformal approach, applied to \eqref{eq:flatmagnGalscalar}, leads to the novel Galilean invariant action \eqref{eq:Lmagnfix}, it does not reproduce the magnetic Galilei gravity action \eqref{magnGalgrav}. This concludes our discussion of the Galilean case.

\section{Discussion}

In this paper we applied the conformal technique to a single Carroll scalar and showed its relation, upon gauge-fixing, with Carroll gravity. In particular, we showed that a massless electric scalar leads to a particular form of electric Carroll gravity while a massless magnetic scalar leads to magnetic Carroll gravity. On the way we showed that  other gravitational  invariants involving the intrinsic torsion tensor exist as well but they were not related to dynamical matter. In the Galilei case we found other examples where the conformal program did not work as in general relativity, in the sense that one can couple Galilean scalars to Galilean conformal gravity but not by following the rules of the conformal Galilei algebra in the same way that we did for the conformal Carroll algebra. This already shows from the  mere fact that in the Galilei case we cannot construct dependent gauge fields for the Galilei special conformal transformations, a feature that played a crucial role in the Carroll calculation. Another peculiar feature of Galilei gravity invariants is that some of them  are invariant under local an-isotopic dilatations.

Starting from the results of this work one may obtain non-trivial couplings of Carroll gravity  to matter  by  replacing the single scalar we have been using in this work  by a function of $N$ scalars of which only one is used for gauge-fixing the local dilatations. One could also consider Carroll gravity coupled to fermions like in \cite{Bergshoeff:2023vfd}.

There are several generalizations that come to mind. First of all, the conformal technique has been heavily applied with great success in the supergravity literature \cite{Freedman:2012zz}. One may extend the results of this paper to the supersymmetric case and construct first examples of a Carroll supergravity theory. Another natural generalization is to consider conformal Carroll algebras with an-isotropic dilatations such as the ones considered in \cite{Afshar:2024llh}. These algebras lead to a different scaling weight for the longitudinal and transverse Vierbeine.
This suggests a realization of conformal Carroll symmetries in terms of a scalar field theory with different powers of time and spatial  derivatives.  Such Lagrangians have appeared in the study of fractons \cite{Pretko:2020cko,Bidussi:2021nmp,Hartong:2024hvs} and spacetime subsymmetries \cite{Baig:2023yaz,Kasikci:2023tvs}.
Yet another generalization is to consider extended objects instead of particles alone. For a $p$-brane, this requires a more general decomposition of the relativistic flat index into $p+1$ longitudinal and $D-p-1$ transverse directions. One could then also consider the geometry underlying Carroll strings \cite{Cardona:2016ytk,Harksen:2024bnh}.

Finally, the Carroll and Galilei algebras are just two examples  of a non-Lorentzian algebra. One could apply the conformal technique to other non-Lorentzian algebras as well. For the Schr\"odinger algebra, this has already been done in \cite{Afshar:2015aku}. Another interesting option, with a view on fractons \cite{Bidussi:2021nmp,Hartong:2024hvs}, is to consider the  Aristotelian algebra which is an intersection of the  Carroll and Galilei algebra breaking the boost symmetry. It would be interesting to understand, for the case of non-Lorentzian algebras, when  the conformal program works and when not. We hope to come back to some of these issues and generalizations soon.


\section*{Acknowledgements}

E.B.~would like to thank the physics department of the Universidad Cat\'olica de la Sant\'\i sima Concepci\'on, Chile
for hospitality and Fondecyt grants No. 11220328 for financial support. He also wishes to thank the organizers of the UTA/UCSC/UDEC/UTALCA 2024: {\sl 1st joint School and Workshop on theoretical gravity and related topics} for providing a stimulating atmosphere. E.B.~and J.R.~would like to acknowledge their participation to the ESI programme {\sl Carrollian Physics and Holography} at the  Erwin Schr\"odinger Institute in Vienna where part of this work was done.
P.C. acknowledges financial support from the National Agency for Research and Development (ANID) through Fondecyt grants No. 1211077 and 11220328. E.R. acknowledges financial support from ANID through SIA grant No. SA77210097 and Fondecyt grant No. 11220486. P.C. and E.R. would like to thank to the Direcci\'{o}n de Investigaci\'{o}n and Vicerector\'{\i}a de Investigaci\'{o}n of the Universidad Cat\'{o}lica de la Sant\'{\i}sima Concepci\'{o}n, Chile, for their constant support. This work is supported by the Croatian Science Foundation project IP-2022-10-5980 “Non-relativistic supergravity and applications”. J.R.~would like to thank the Erwin Schr\"odinger Institute in Vienna for hospitality and support via its "Research in Teams" programme, during part of this work.


\providecommand{\href}[2]{#2}\begingroup\raggedright\endgroup

\end{document}